\documentclass[aip,jcp,
11pt,
amsmath,amssymb,
floatfix,
reprint,
]{revtex4-1} 

\usepackage[utf8]{inputenc} 
\usepackage[T1]{fontenc} 
\usepackage{hyperref}
\usepackage{verbatim}
\usepackage[caption=false]{subfig}

\usepackage[table]{xcolor} 
\usepackage{longtable}
\usepackage{graphicx}

\usepackage{booktabs}
\usepackage{dcolumn}\usepackage{bm}

\usepackage{listings}

\usepackage[version=3]{mhchem}
\usepackage{units}
\usepackage{siunitx}

\usepackage{xspace}
\usepackage[textsize=small]{todonotes}

\graphicspath{{pics/}{pics/decay/}}

\newcommand{\matr}[1]{\textbf{#1}}

\newcommand{\braket}[2]{\ensuremath{ \langle #1 | \, #2  \rangle }}
\newcommand{\ketbra}[2]{\ensuremath{  | {#1} \rangle \langle {#2} |}}
\newcommand{\ket}[1]{\ensuremath{  | {#1} \rangle}}
\newcommand{\bra}[1]{\ensuremath{\langle {#1} | }}
\newcommand{\matrixe}[3]{\ensuremath{ \langle{#1} | \vphantom
        {#1 #3} {#2}
| {#3} \rangle }}

\definecolor{ocre}{RGB}{243,102,25}
\definecolor{mygray}{RGB}{243,243,244}
\definecolor{fzjred}{RGB}{175,90,80}

\definecolor{blau}{HTML}{1F78B4}
\definecolor{gruen}{HTML}{33A02C}
\definecolor{hellblau}{HTML}{A6CEE3}
\definecolor{hellgruen}{HTML}{B2DF8A}

\definecolor{nrot}{HTML}{d7191c}
\definecolor{norange}{RGB}{253,174,97}
\definecolor{ngruen}{HTML}{abdda4}
\definecolor{nblau}{HTML}{2b83ba}

\definecolor{nrot1}{RGB}{215,48,31}
\definecolor{nrot2}{RGB}{252,141,89}
\definecolor{nrot3}{RGB}{253,204,138}
\definecolor{nrot4}{RGB}{254,240,217}

\definecolor{CBred}{RGB}{215,25,28}
\definecolor{CBorange}{RGB}{253,174,97}
\definecolor{CByellow}{RGB}{255,255,191}
\definecolor{CBgreen}{RGB}{171,211,164}
\definecolor{CBlgreen}{RGB}{166,217,106}
\definecolor{CBdgreen}{RGB}{26,150,65}
\definecolor{CBblue}{RGB}{43,131,186}
\definecolor{CBblue2}{RGB}{146,197,222}
\definecolor{CBdblue}{RGB}{5,113,176}
\definecolor{CBgray60}{RGB}{102,102,102}
\definecolor{CBgray20}{RGB}{204,204,204}

\newcommand{\newi}{}

\let\oldtheequation\theequation
\makeatletter
\def\tagform@#1{\maketag@@@{\ignorespaces#1\unskip\@@italiccorr}}
\renewcommand{\theequation}{(\oldtheequation)}
\makeatother

\begin{document}

\title{Computing vibrational eigenstates with tree tensor network states (TTNS)}

\author{Henrik R.~Larsson}
\email{larsson [a t] caltech . e$\delta$u}
\affiliation{Division of Chemistry and Chemical Engineering, California Institute of Technology, Pasadena, CA 91125, USA}
\date{\today}

\begin{abstract}
We present how to compute vibrational eigenstates with tree tensor network states (TTNS), the underlying 
\emph{ansatz} behind the multilayer multi-configuration time-dependent Hartree (ML-MCTDH) method.
The eigenstates are computed with an algorithm that is based on the density matrix renormalization group (DMRG). We apply this to compute the vibrational spectrum of acetonitrile (\ce{CH3CN}) to high accuracy and compare TTNS with matrix product states (MPS), the \emph{ansatz} behind the DMRG.
\newi{The presented optimization scheme converges much faster than ML-MCTDH-based optimization.}
For this \newi{particular} system, we found no major advantage of the more general TTNS over MPS. 
We highlight that for both TTNS and MPS, the usage of an adaptive bond dimension significantly reduces the amount of required parameters.
We furthermore propose a procedure to find good trees.
\end{abstract}

\maketitle

\section{Introduction}
\label{sec:intro}

The multi-configuration time-dependent Hartree (MCTDH) method has been established as a very powerful method for performing molecular quantum dynamics simulations.\cite{mctdh_cederbaum_1990,mctdh_NOCl_cederbaum_1992,mctdh_rev_meyer_2000,mctdh_book} 
Mathematically, the MCTDH \emph{ansatz} is based on a tensor decomposition, namely the Tucker decomposition.\cite{tensor_decomp_rev_bader_2009}
In its standard form, it can be applied to up to 12-dimensional systems.\cite{Huarte-Larranaga2001}
With mode-combination, that is, a redistribution of dimensions within the tensor decomposition, it has been successfully applied to up to about 80-dimensional model systems.\cite{pyrazine_24d_cederbaum_1998,wang_2000,nest_2003}
However, MCTDH inherently scales exponentially and cannot be applied to larger and more complex systems.
Major improvements could be obtained with a generalization of MCTDH, namely the multilayer MCTDH (ML-MCTDH) method.\cite{ml_mctdh_thoss_2003,ml_mctdh_manthe_2008,ml_mctdh_meyer_2011,ml_mctdh_rev_wang_2015,manthe_2017}
In mathematics, this approach is known as hierarchical Tucker decomposition\cite{hierarchical_tucker_decomp_kuehn_2009,hierarchical_tucker_decomp_grasedyck_2010,hackbusch_book}
whereas in physics, it is known as tree tensor network states (TTNS) decomposition,\cite{shi_2006a,tagliacozzo_2009,murg_2010} a subset of the more general tensor network states (TNS) decomposition.\cite{Chan2012,Orus2014a} 
With ML-MCTDH, scientists were able to simulate model systems with more than 1000 dimensions.\cite{ml_mctdh_thoss_2003,ml_mctdh_meyer_2011,ml_mctdh_anthracene_c60_lan_2015,ml_mctdh_FMO_kuehn_2016,mendive-tapia_2018}

Independently, related tensor decompositions have also been developed in condensed matter physics, electronic structure theory and other fields.\cite{Chan2012,Orus2014,Orus2014a,Schollwoeck2011}
In particularly successful in these fields is the density matrix renormalization group (DMRG).\cite{white_1992,white_1993,Schollwoeck2011}
The DMRG is based on a so-called matrix-product state (MPS) tensor decomposition.\cite{Schollwoeck2011}
This is kown in mathematics as tensor train (TT) decomposition.\cite{oseledets_2011a}
While MPS actually are a subset of TTNS and thus very much related to ML-MCTDH, the DMRG uses a completely different approach for performing quantum simulations.
ML-MCTDH is based on the Dirac-Frenkel time-dependent variational principle\cite{mctdh_rev_meyer_2000} and gives highly nonlinear  equations of motions whereas the DMRG, essentially, converts the nonlinear (typically time-independent) equations to fixed-point iterations, similar to the Hartree-Fock \newi{self-consistent field } algorithm.\cite{chan_dmrg_lagrangians_2008}

There has not been much overlap between ML-MCTDH and DMRG developments.
Only recently, developments for MPS and for the DMRG have been transferred to molecular quantum dynamics.\cite{Rakhuba2016a,greene_2017,Baiardi2017,chan_2018,kurashige_2018,baiardi_2019,baiardi_2019a,schroder_2019,iouchtchenko_2018}
However, much remains to be done in order to systematically compare and apply the different approaches to molecular quantum dynamics.
In particular, we are not aware of any \emph{direct} comparisons between MPS and ML-MCTDH/TTNS for molecular systems.

This paper is aimed to make a step into this direction. The aim of this paper is threefold: (1) We present and apply the highly successful diagrammatic notation\cite{shi_2006a,tagliacozzo_2009,Orus2014a}
in the context of molecular quantum dynamics. 
This powerful notation is very common in physics,
but has, so far, been used in molecular quantum dynamics only as a pictorial tool\cite{ml_mctdh_manthe_2008,thomas_hcp_2015} and not for deriving equations.
(2) We use this notation to transfer the essentials of the DMRG algorithms to TTNS in order to be able to compute vibrational spectra (i.e., to solve the time-independent Schrödinger equation). We also focus on using adaptive tensor sizes throughout the simulations and how to find good trees.
(3) With the same methodology and code, we directly compare MPS with TTNS for the 12-dimensional \ce{CH3CN} molecule.

To distinguish our methodology from ML-MCTDH, we use the term TTNS for our work in the following, even though ML-MCTDH is based on TTNS for solving the time-dependent Schrödinger equation.
Both TTNS and ML-MCTDH thus have the same overall computational scaling. The only difference is the way to apply tensor decompositions to quantum systems.

\newi{With ML-MCTDH, vibrational spectra have been computed  
using combinations of time-independent Krylov subspace techniques and 
imaginary time evolution based on the time-dependent variational principle.}\cite{hammer_2012,wodraszka_2012,wang_2014c}
\newi{The nonlinear nature of the ML-MCTDH equations of motions makes
it} often difficult to use it for computing many eigenstates. 
In contrast, since the very beginning of TTNS,\cite{tagliacozzo_2009,murg_2010,changlani_2013,nakatani_2013,gerster_2014,gunst_2018} 
they have been used with DMRG algorithms that directly solve the time-independent Schrödinger equation \newi{ by iteratively solving many quadratic (eigenvalue) problems}. This should allow for a more direct and straightforward calculation of vibrational spectra with many high-lying excited states.

This work thus aims to complement the already well-established ML-MCTDH method and to give an alternative approach to compute spectra. 
Note that approaches based on related tensor decompositions (the Candecomp format) already provide another alternative.\cite{Leclerc2014,thomas_hcp_2015,thomas_2017,thomas_2018}

The remainder of this Article is organized as follows: \autoref{sec:theory} gives a detailed overview of the used tensor decompositions and tensor network states, the diagrammatic notation to present tensor network states and DMRG-like optimizations for TTNS. \autoref{sec:appl_ch3cn} shows an application of TTNS to the computation of the vibrational spectrum of acetonitrile, \ce{CH3CN}, and compares \newi{both} TTNS with MPS \newi{and the ML-MCTDH-based optimization with the DMRG-like optimization}. \autoref{sec:conclusion} concludes and gives an outlook.

\section{Theory}
\label{sec:theory}

In the following, the general notation and theory is discussed. After an introduction of tensor decompositions, tensor network states and the diagrammatic notation in \autoref{sec:theory_intro}, the exploitation of gauge degrees of freedom are discussed in \autoref{sec:theory_canonicalization}. 
\autoref{sec:theory_ground_state} discusses the used algorithm for ground state minimization and \autoref{sec:theory_exited_states} discusses how to obtain excited states. This Section ends with a description of how to adapt the number of parameters in the tensor networks in \autoref{sec:theory_bond_adaption} and how to find good trees in \autoref{sec:theory_disentangling}. 
Some of the used notation and symbols are summarized in \autoref{tab:notation}.
Here, we mostly (but not exclusively) use the notation from the ML-MCTDH context.\cite{ml_mctdh_manthe_2008,ml_mctdh_meyer_2011,Otto2014}

\begin{table}
  \caption{Some of the notation used in this work. A non-exhaustive list of alternative symbols are also given. ``\#'' stands for ``number.''
  \label{tab:notation}
  }
 \begin{ruledtabular}
 \begin{tabular}{lll}
   description & symbol & alternative symbols\\ \hline
physical dimension               & $F$          & $D$\\
physical index                   & $f$                & $\kappa$\\
physical basis                   & $\ket{\chi^{(f)}}$ & $\ket{\sigma}$\\
physical basis dimension         & $N^{[f]}$          & $n_f$, $k$ \\
\# of layers & $L$ \\
layer        & $l$ \\
horizontal position in layer $l$ & $\kappa$           & $n$ \\
bond dim./rank/\# of SPFs   & $n^{[l;\kappa]}$   & $m$, $M$, $D$, $\chi$, $r$\\
dimension of node/tensor/site    & $d^{[l;\kappa]}$   & \\
node/SPF tensor/site             & $A^{[l;\kappa]}$   & $\chi^{[l,\kappa]}$, $\Lambda^{[l,\kappa]}$\\
\end{tabular}
 \end{ruledtabular}
\end{table}

\subsection{Tensor decompositions}
\label{sec:theory_intro}
To solve Schrödinger's equation, the $F$-dimensional wavefunction typically is represented
\emph{via} a Galerkin approach.
That is, a direct product of (often orthogonal) ``primitive'' bases $\{\ket{\chi^{(f)}_{\alpha_f}}\}_{\alpha_f=1}^{N_f}$ of finite size $N_f$ for dimension $f\in[1,F]$ is used as representation:
\begin{equation}
  \ket{\Psi} = \sum_{\alpha_1}^{N_1}  \sum_{\alpha_2}^{N_2} \dots \sum_{\alpha_F}^{N_F} C_{\alpha_1 \alpha_2\dots \alpha_F} \bigotimes_{f}^{F} \ket{\chi^{(f)}_{\alpha_f}}.
  \label{eq:fci}
\end{equation}
The entries of the real- or complex-valued coefficient tensor $\matr C$ are then to be determined.

While this approach is very simple and numerically robust, the problem lies in the 
the coefficient tensor $\matr C$ whose size scales exponentially as $N_\text{mean}^{F}$ where $N_\text{mean}$ is the geometric mean of the number of employed basis functions per dimension.
For molecular quantum dynamics, $N_\text{mean}$ typically is $\sim 10$ and this approach is then limited to about $F=9$ dimensional systems. Tensor decompositions lower the scaling and enable studies for much larger systems.

For introducing tensor decompositions, specifically tensor networks, we consider a $F=3$ dimensional wavefunction represented in a finite basis as shown in \autoref{eq:fci}. The coefficient tensor then  has entries  $C_{\alpha \beta \gamma}$. 
In the so-called Tucker decomposition, which is the underlying decomposition of the MCTDH approach,
$\matr C$ is factorized in an up to $F$-dimensional so-called core-tensor $\matr A^{[1]}$ and several smaller-dimensional, ``auxiliary'' tensors $\matr A^{[2,\kappa]}$:
\begin{equation}
  C_{\alpha \beta \gamma} \approx \sum_{ijk} A^{[1]}_{ijk} A^{[2,1]}_{\alpha i} A^{[2,2]}_{\beta j} A^{[2,3]}_{\gamma k},\label{eq:tucker}
\end{equation}
where $i \in [1,n^{[2,1]}]$, $j \in [1,n^{[2,2]}]$ and $k \in [1,n^{[2,3]}]$.
Here, $n^{[l,\kappa]}$  are called  ``bond dimensions'', ``ranks'' or ``number of single-particle functions (SPFs)'' for a particular tensor $\matr A^{[l,\kappa]}$. 
We will use the term bond dimension in the following.
In $n^{[l,\kappa]}$ and $\matr A^{[l,\kappa]}$,
$l$ is the \emph{layer} and $\kappa$ the horizontal position in a particular layer.
The Tucker decomposition thus is a two-layer approach. The notion of layers will become more clear shortly.
Note that we use here a more elaborate notation because this will be required for the following more intricate tensor decompositions.

Since the core tensor $\matr A^{[1]}$ has the same dimensionality as the original coefficient tensor $\matr C$, the Tucker decomposition does not avoid exponential scaling with dimensionality.
Nevertheless, the problem size typically is reduced due to the introduction of the auxiliary tensors $\matr A^{[2,\kappa]}$, which are contracted with the core tensor in order to restore $\matr C$.
For many problems, the required bond dimensions $n^{[l,\kappa]}$  are much smaller than the physical dimensions $N^{[f]}$ of the coefficient tensor. This drastically reduces the size of the core tensor, compared to the coefficient tensor.
Despite this reduction of problem size, the exponential scaling of the core tensor typically limits the Tucker decomposition to about \newi{$15$-dimensional systems,\cite{Huarte-Larranaga2001,vendrell_2007c}} or, for model systems, to about $80$-dimensional systems.\cite{pyrazine_24d_cederbaum_1998,wang_2000,nest_2003}

The exponential scaling of the core tensor can be avoided by 
lowering its dimensionality. This is achieved by introducing higher-dimensional auxiliary tensors. 
They are then decomposed further using, again, a Tucker decomposition; for example,
\begin{align}
  C_{\alpha \beta\gamma} &\approx \sum_{ij} A^{[1]}_{ij} A^{[2,1]}_{\alpha i} B_{\beta \gamma j}\\
   &= \sum_{ij} A^{[1]}_{ij} A^{[2,1]}_{\alpha i} \sum_{kl} A^{[2,2]}_{klj} A^{[3,1]}_{\beta k} A^{[3,2]}_{\gamma l}\label{eq:ttns}
\end{align}
Here, the core tensor $\matr A^{[1]}$ is only two-dimensional and the higher-dimensional auxiliary tensor $\matr B$ is again Tucker-decomposed into $\matr A^{[2,2]}$, $\matr A^{[3,1]}$ and $\matr A^{[3,2]}$.\footnote{Not decomposing the tensor $\matr B$ leads to the so-called mode-combination approach in standard MCTDH.\cite{pyrazine_24d_cederbaum_1998} It allows for simulating larger systems but is still inherently limited with respect to dimensionality, even though improvements are possible.\cite{dpmctdh_tannor_2017}}
This type of decomposition is the underpinning of the ML-MCTDH method and of TTNS. Note that in many TTNS calculations in electronic structure theory,\cite{nakatani_2013,murg_2015,gunst_2018}
(almost) every tensor in the TTNS typically is connected with a physical dimension whereas in vibrational dynamics, only the tensors in the lowest layers typically are connected with a physical dimension.\cite{ml_mctdh_rev_wang_2015}
This difference, however, is not fundamental but rather depends on the best way to represent the physical system (Fermions with a ``generic'' Hamiltonian vs.~distinguishable particles with a Hamiltonian that leads to some approximate grouping of particles, in terms of the coupling).

As a special case of TTNS, one can formulate a tensor decomposition where each value of $C_{\alpha\beta\gamma}$ is reconstructed by computing a trace of product of matrices:
\begin{equation}
  C_{\alpha \beta\gamma} \approx \sum_{i} A^{[1]}_{\alpha i} \sum_j A^{[2]}_{\beta ij} A^{[3]}_{\gamma j}. \label{eq:mps}
\end{equation}
The trace of matrix products can be understood by considering $\matr A^{[2]}_{\beta ::}$ as a matrix and $\matr A^{[1]}_{\alpha :}$ ($\matr A^{[3]}_{\gamma :}$) as row (column) vectors.
Hence, this decomposition is called matrix product state (MPS) and is the \emph{ansatz} behind the 
DMRG.

\begin{figure}
  \includegraphics{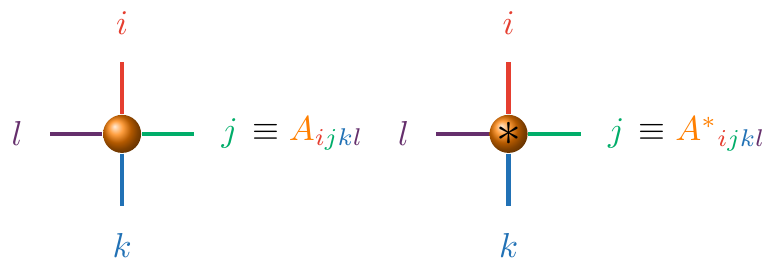}
  \includegraphics{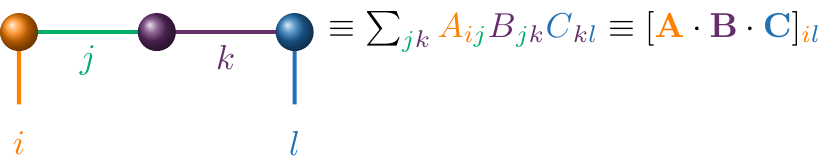}
  \caption{Diagrammatic notation used in this work. The upper panel shows the four-dimensional tensor $A_{ijkl}$ (left) and its complex conjugate (right) represented in diagrammatic notation. The lower panel shows a particular tensor contraction, the $il$th entry of the matrix product $\matr A \matr B \matr C$. 
  See text for further details.}
  \label{fig:dia_notation}
\end{figure}

Since it is quite tedious to deal with these decompositions using equations, here, we instead use
a diagrammatic notation.\cite{Chan2012,Orus2014a}
\autoref{fig:dia_notation} gives an overview of it.
There, each node represents a tensor.
A node with an asterisk marks complex conjugation.\footnote{While here we only use real-valued tensors, we hint at complex conjugation in order to be general and to distinguish whether $\ket{\Psi}$ or its dual $\bra{\Psi}$ is represented in a particular diagram.} 
In the following, we use the terms node and tensor interchangeably. In the context of DMRG, nodes are called ``sites.''
Each vertex represents a tensor dimension. Vertices that connect two nodes/tensors represent a contraction, that is, a summation between the two tensors over the particular dimension.

\begin{figure}
  \includegraphics{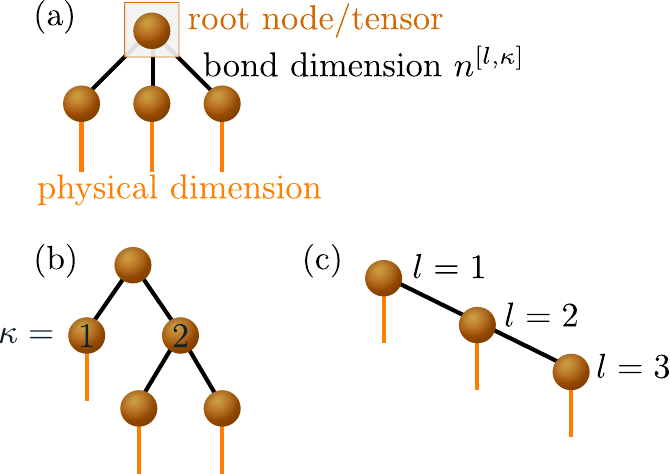}
  \caption{Examples of tensor network states/decompositions in diagrammatic notation. 
    (a) shows the Tucker decomposition from \autoref{eq:tucker}, (b) shows the tree tensor network state (TTNS) from \autoref{eq:ttns} and (c) shows the matrix product state (MPS) from \autoref{eq:mps}.
    The symbols for the bond dimension $n^{[l,\kappa]}$ (shown in A) in layer $l$ (shown in (c)) and horizontal position $\kappa$ (shown in (b)) are also exemplified.}
  \label{fig:tensor_decomps}
\end{figure}

The previously introduced examples of tensor network states are shown in  \autoref{fig:tensor_decomps} using the diagrammatic notation. 
There, each tensor $\matr A^{[l,\kappa]}$ is labeled by layer $l$ and horizontal position $\kappa$ in the tensor network. 
The dangling bonds represent the physical dimensions of size $N^{[f]}$.
The vertices that connect tensors introduce ``virtual'' bond dimensions of size $n^{[l,\kappa]}$.
Contracting all virtual bonds finally restores the core tensor $\matr C$. 
The notion of root tensor will become clear in the following \autoref{sec:theory_canonicalization}.

Another notation is to associate each tensor $\matr A^{[l,\kappa]}$ with the representation of some basis function.
In the context of DMRG, these functions are called renormalized basis functions whereas in the context of MCTDH, they are known as single-particle functions (SPFs), $\ket{\phi}$.
SPFs can be understood as a variationally optimized basis and are represented either by SPFs in higher layers or by the primitive basis, e.g., $\ket{\phi^{[l,\kappa]}_{i}} = \sum_{\alpha} A^{[l+1,\kappa_{l+1}]}_{\alpha i} \ket{\chi^{(\kappa_{l+1})}_\alpha}$.
While this basis notation is very powerful as well and can go hand-in-hand with the diagrammatic notation, here, we mostly use the latter. However, for some cases, the former notation is simpler and hence will be used occasionally. With SPFs, the wavefunction from the three-dimensional example can be described in terms of multidimensional configurations $\ket{\Phi_{ijk}}$ that are decomposed as product of SPFs:
\begin{align}
\ket{\Psi} &= \sum_{ijk} A^{[1]}_{ijk} \ket{\Phi^{[1]}_{ijk}}\label{eq:spf_def} = \sum_{ijk}  A^{[1]}_{ijk} \ket{\phi^{[1,1]}_{i}} \ket{\phi^{[1,2]}_{j}} \ket{\phi^{[1,3]}_{k}} \\
            &= \sum_{\alpha \beta \gamma} \sum_{i j k} A^{[1]}_{ijk}  A^{[2,1]}_{\alpha i} A^{[2,2]}_{\beta j} A^{[2,3]}_{\gamma k}   \ket{\chi^{(1)}_{\alpha}} \ket{\chi^{(2)}_{\beta}} \ket{\chi^{(3)}_{\gamma}}.
\end{align}

\subsection{Canonicalization}
\label{sec:theory_canonicalization}

\begin{figure}
     \includegraphics{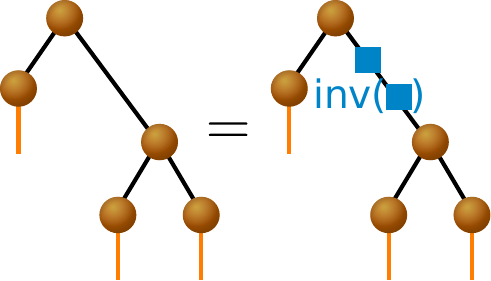}
     \caption{Example of the gauge degree of freedom. A matrix (shown as blue rectangle) and its inverse is inserted into a particular bond. The matrix and its inverse can be absorbed into the neighboring tensors without changing the final, contracted tensor.}
     \label{fig:gauge}
\end{figure}

\begin{figure}
     \includegraphics{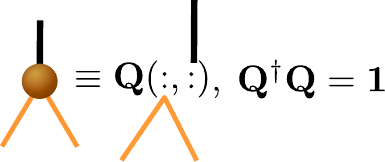}
     \caption{Representation of a tensor as orthogonal matrix (matricization). The two lower bonds represent the rows (as multi-index) of the matrix $\matr Q$ and the upper bond represent the columns.}
     \label{fig:tens_orth_mat}
\end{figure}

\begin{figure}
     \includegraphics{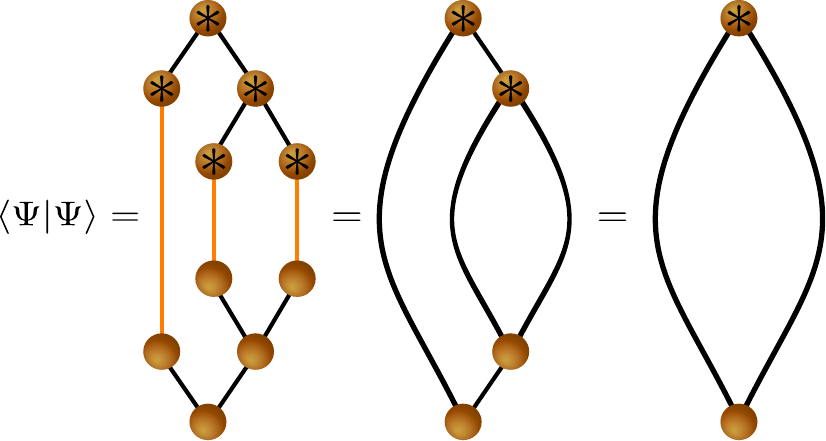}
     \caption{Norm calculation for a tensor network state with orthogonalized tensors. The contraction of the tensor network simplifies due to the orthogonality of each tensor; compare with \autoref{fig:tens_orth_mat}.}
     \label{fig:Bracket}
\end{figure}

The presented tensor networks actually have some intrinsic redundancy:
An insertion of an invertible matrix and its inverse between each bond does not change the overall
wave function; see \autoref{fig:gauge}.
This so-called gauge degree of freedom can be exploited to improve numerics.\cite{mctdh_rev_meyer_2000,Schollwoeck2011} In this Section, we discuss two ways to improve numerics and will later in \autoref{sec:theory_sweep} discuss a third way.
The first way is to reshape each tensor into a matrix and restrict this matricized\cite{tensor_decomp_rev_bader_2009}
tensor to be orthogonal, c.f.~\autoref{fig:tens_orth_mat}.
In the used tensor diagrams, the matricization is implicitly declared by the location of the bonds on each tensor. 
Bonds pointing away from the root node (typically downwards)  define a multi-index that is associated with the rows of the matrix. The bond that points towards the root node is associated with the columns of the matrix. 
\footnote{Another frequently used notation is to use arrows or triangles to specify the matricization.\cite{Orus2014a,haegeman_2016}} 
Note that in all tensor networks shown here, for each node, there is only one bond pointing to the root node (meaning that the tensor network does not contain loops).
The orthogonalization of the matricized tensors means nothing else than that the SPFs are orthogonal.
It does massively simplify many expressions. For example, the calculation of the norm, $\sqrt{\braket{\Psi}{\Psi}}$, boils down to the norm calculation of the root tensor (viewed as a vector). This is shown in  \autoref{fig:Bracket}.

\begin{figure}
     \includegraphics{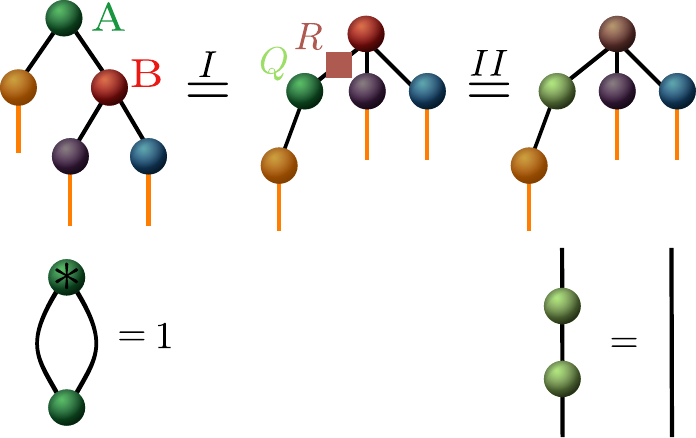}
     \caption{Change of canonicalization (root node) from tensor $\matr A$ to tensor $\matr B$. The upper panel show the steps performed in the canonicalization: Reshaping and QR decomposition in step $I$ and absorption of the $\matr R$ matrix in step $II$. The lower panels show the particular orthogonality conditions (isometries) of tensor $\matr A$ for the canonical forms where $\matr A$ is root node (left) and where $\matr B$ is root node (right). A single line represents a unit matrix.}
     \label{fig:QR}
\end{figure}

The orthogonalization towards a specific (root/central) node is called canonicalization. A change of the canonical form (change of the root node) is possible \emph{via} a QR
or similar matrix decomposition,\cite{golub_book} see \autoref{fig:QR}: Starting from the previous root node, the tensor is matricized such that the bond pointing toward the new root node represents the columns (step $I$ in \autoref{fig:QR}).
Then, a QR decomposition is performed: $\matr A = \matr Q \matr R$. The orthogonal matrix $\matr Q$
becomes the new tensor and the triangular matrix $\matr R$ is absorbed into the neighbor (step $II$ in \autoref{fig:QR}).
This procedure is repeated until the desired new root node is reached.

While the canonicalization makes every (matricized) tensor orthogonal, any unitary transformation
between two connected tensors is still possible. Thus, a second way to exploit the redundancy is to 
fix this unitary transformation.  
In the context of MCTDH, this unitary degree of freedom is fixed by adding an additional gauge 
operator to the Hamiltonian. Some gauges simplify the solution of the differential equations.\cite{mctdh_rev_meyer_2000,madsen_2018a}
In contrast to MCTDH, here, this gauge is \emph{not} fixed directly.
Finding most suitable gauges is subject to future work. 
In the context of MCTDH and improved relaxation, so-called energy orbitals are used to fix the gauge.
They diagonalize the separable part of the Hamiltonian.\cite{doriol_2008}

An alternative is to use natural orbitals that diagonalize reduced density matrices. 
While natural orbitals are not used directly in this work, changing the canonical form (during the sweeps to be introduced in \autoref{sec:theory_sweep}) actually leads to this gauge for the root node.

\subsection{Ground state optimization}
\label{sec:theory_ground_state}

Having introduced TTNS, we now discuss how to optimize for the ground state. That is, the 
goal is to minimize the expectation value of the Hamiltonian $\hat H$:
\begin{equation}
  E = \text{min}_\Psi \frac{\matrixe{\Psi}{\hat H}{\Psi}}{\braket{\Psi}{\Psi}}
\end{equation}
For that, we first discuss possible forms of $\hat H$ in \autoref{sec:theory_hamilt} and then discuss the employed sweep algorithm in \autoref{sec:theory_sweep}.

\subsubsection{Hamiltonian}
\label{sec:theory_hamilt}

In the primitive basis $\{\ket{\chi^{(f)}_{\alpha_f}}\}_{\alpha_f=1}^{N_f}$ (see \autoref{eq:fci}), the Hamiltonian $\hat H$ has the following matrix representation:
\begin{align}
H_{\alpha_1\alpha_2\dots \alpha_F, \alpha'_1\alpha'_2\dots \alpha'_F} &= \bigotimes_{f=1}^F\bigotimes_{f'=1}^F \matrixe{\chi_{\alpha_f}^{(f)}}{\hat H}{\chi_{{\alpha'}_{f'}}^{(f')}},
\end{align}

For methods based on tensor decompositions of $\Psi$, working equations become much simplified if $\hat H$ takes a similar (not necessarily identical) tensor decomposition as $\Psi$ itself.
Hence, we assume here that $\hat H$ can be decomposed as a sum of direct products of one-dimensional operators or matrices in finite basis representation (SOP):
\begin{align}
H_{\alpha_1\alpha_2\dots \alpha_F, \alpha'_1\alpha'_2\dots \alpha'_F} &\approx
\sum_{s=1}^{N_\text{PF}} c_s \bigotimes_{f=1}^F  h^{(f,s)}_{\alpha_f,\alpha_f'},\label{eq:sop_structure}
\end{align}

\begin{figure}
    \includegraphics{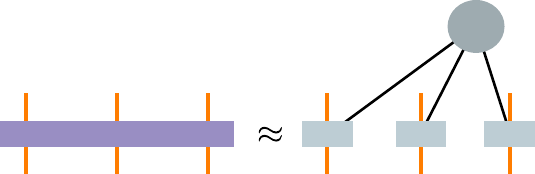}
     \caption{Tucker decomposition (``potfit'') of the Hamiltonian for a three-dimensional example (this leads to a six-dimensional Hamiltonian tensor to be decomposed).}
     \label{fig:H_decomp}
\end{figure}

Such a decomposition can be achieved, for example, using the ``potfit'' algorithm\cite{potfit_meyer_1996}
or variants thereof\cite{multigrid_potfit_meyer_2013,schroder_2017,Otto2014,pes_neural_network_sum_of_products_carrington_2006,sop_pes_neural_networks_zhang_2014}
that Tucker-decompose the potential in grid representation.\footnote{The kinetic energy operator very often is already in SOP form.}
The potfit procedure is shown diagrammatically in \autoref{fig:H_decomp}.
Another alternative is the related Candecomp format.\cite{tensor_decomp_rev_bader_2009}
There, the circle-shaped tensor in \autoref{fig:H_decomp} would be diagonal.
Most optimal decompositions would probably be decompositions of the Hamiltonian using similar tree tensor networks (tree tensor network operators). This is called multilayer potfit for MCTDH.\cite{Otto2014} 
It has previously been used for the special case of matrix products/tensor trees by Rakhuba and Oseledets.\cite{Rakhuba2016a}
Other kinds of approximations to avoid the tensor decomposition at all may also be possible in certain cases.\cite{ml_mctdh_manthe_2008,wodraszka_2018}
For many-body decompositions of the Hamiltonian, so-called complementary operators can also be used.\cite{xiang_1996}
While all of this would certainly decrease the overall computational effort we here stick to the standard SOP format.

\subsubsection{Sweep algorithm}
\label{sec:theory_sweep}

We now address the actual ground state minimization. 
The tensor decomposition turns the eigenvalue problem into a nonlinear optimization problem. However, the decomposition also clearly identifies ways to approximately decouple the tensors and to optimize the tensor network tensor by tensor, thereby solving many smaller subproblems instead of solving one big nonlinear problem.
This idea has already been used in the context of improved relaxation in MCTDH,\cite{Meyer2006,doriol_2008}
where the root node is solved separately from the other nodes (see also Refs.~\citenum{Culot1994,drukker_1997,Heislbetz2010}). 

\begin{figure}
     \includegraphics{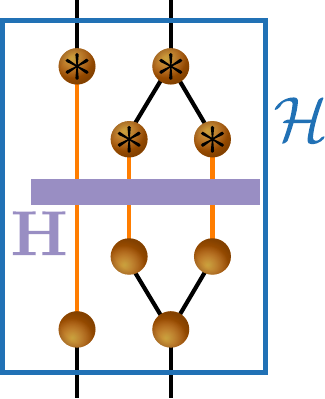}
     \caption{Effective Hamiltonian $\mathcal H$ from \autoref{eq:MF} (blue rectangle) in diagrammatic notation. The purple block represents the full Hamiltonian $\matr H$ (see \autoref{fig:H_decomp} for simplifications).
     The three-dimensional wavefunction  is represented by the tensor network (b) shown in \autoref{fig:tensor_decomps}.}
     \label{fig:MF}
\end{figure}

Keeping all tensors but the root tensor fixed results in a standard eigenvalue problem for the optimal values of the root tensor\cite{drukker_1997,Meyer2006}
with an effective Hamiltonian $\mathcal H$ obtained by representing the full Hamiltonian in configuration representation (c.f.~\autoref{eq:spf_def}):
\begin{equation}
\mathcal H = \matrixe{\Phi^{[1]}}{\hat H}{\Phi^{[1]}}\label{eq:MF}
\end{equation}
This is shown diagrammatically in \autoref{fig:MF}.
It can be computed in a recursive way as explained by Manthe.\cite{ml_mctdh_manthe_2008}

In improved relaxation in MCTDH, the minimization of the remaining tensors is then performed via a normal nonlinear optimization for each tensor, typically either by imaginary time evolution\cite{Meyer2006}
or by Jacobi rotations.\cite{drukker_1997,Heislbetz2010} 
This nonlinear optimization is most of the time not the main effort in the optimization for standard MCTDH. However, for ML-MCTDH/TTNS the nonlinear optimization becomes non-trivial, especially for excited states.
\newi{Furthermore, improved relaxation relaxes the SPFs in each degree of freedom independently from each other and thus neglects the coupling between them.}

\begin{figure}
     \includegraphics[width=\columnwidth]{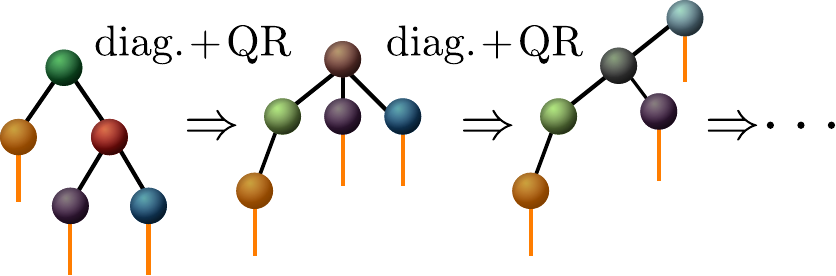}
     \caption{Sweep through a tensor network. At each step, the root node is replaced by the eigenstate of the effective Hamiltonian $\mathcal H$ (\autoref{fig:MF}) and a QR decomposition is performed to change the root node (\autoref{fig:QR}).}
     \label{fig:sweep}
\end{figure}

Here, we perform a different approach,\cite{tagliacozzo_2009,murg_2010,murg_2015,changlani_2013,gerster_2014}
borrowed from the DMRG algorithm.\cite{white_1992,white_1993}
After minimization of the root node while keeping all other nodes fixed, we change the canonical form to another node. This is then minimized by diagonalizing an effective Hamiltonian $\mathcal H$. 
This procedure is repeated for all tensors, resulting in a sweep through the whole tree; see \autoref{fig:sweep}. 
The sweeps are repeated until convergence. 
Thus, the nonlinear optimization problem of minimizing all tensors at once is converted into a fixed-point iteration with  successive quadratic (eigenvalue) problems.

Note that the convergence of the sweep algorithm can be improved by using a two-site variant where two neighboring tensors are first contracted,  then diagonalized simultaneously and afterwards decomposed using a singular value decomposition (SVD).\cite{golub_book}
This leads to faster and more stable convergence but is also more costly. 
Perturbative approaches are also possible to improve convergence of the one-site algorithm.\cite{white_2005,hubig_2015}

\begin{figure}
     \includegraphics{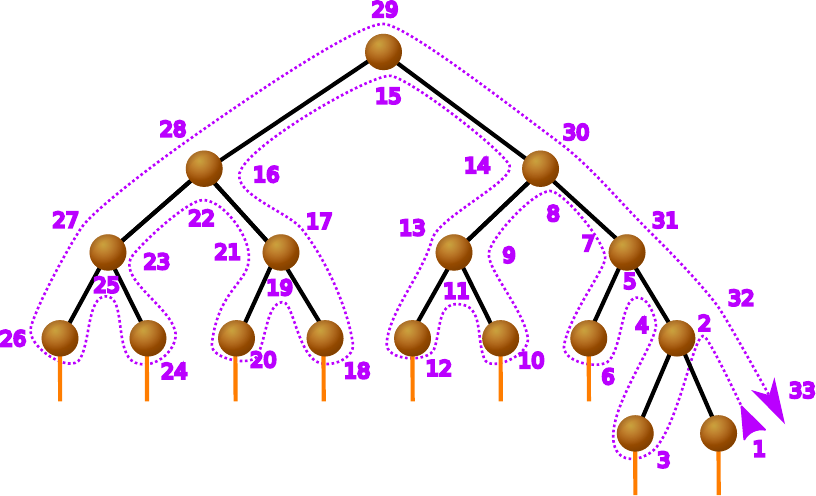}
     \caption{Example of the sweep schedule (purple path; starting in the lower right corner) used in this work. For each number, the corresponding (closest) tensor is optimized.}
     \label{fig:sweep_schedule}
\end{figure}

For MPS, which form linear tensor networks, the way to sweep through the network is clearly identified by sweeping from one end to the other end (forward sweep) and back (backward sweep). For trees, there are several possibilities.
Gerster \emph{et al.} start from the highest layer and optimize the tree layer by layer, finally optimizing the tensors connected to physical dimensions (similar to breadth-first search).\cite{gerster_2014}
Here, we generalize the MPS sweep using a more depth-first search approach such that each tensor is diagonalized as many times as it has bonds per forward and backward sweep: The largest linear chain in the tree is identified and the optimization starts at one end of this chain. One then sweeps through this chain, enters branches whenever they occur and sweeps through the branches forwards and backwards. This procedure is repeated recursively. 
This is very similar to the procedure used by Nakatani and Chan.\cite{nakatani_2013}
An example is shown in \autoref{fig:sweep_schedule}.

We note that the sweep algorithm has overall the same computational scaling as the standard ML-MCTDH algorithm (namely $n^{d+1}$ for the diagonalization with Hamiltonians in SOP form; $n$ is the geometric mean of the bond dimensions on each tensor and $d$ is the dimensionality of the tensor). 
The sweep algorithm thus just presents an alternative way to obtain eigenstates. It may, however, be advantageous as only many eigenvalue problems need to be solved instead of having the necessity to solve nonlinear equations directly.
\newi{The sweep algorithm also takes some coupling into account by allowing, after each update of a tensor, for a feedback to the neighboring tensor to be updated.} 
In contrast to the ML-MCTDH algorithm, no regularization in the differential equations is required.
Even though the regularization in MCTDH can be improved,\cite{mctdh_denmat_inv_lubich_2015,mctdh_denmat_inv_lubich_2017,mctdh_denmat_inv_fischer_2014,mctdh_denmat_inv_manthe_2015,meyer_2018,wang_2018d} here, this is not even necessary.

\subsection{Excited states optimization}
\label{sec:theory_exited_states}

In principle, excited states can be computed using the same procedure as the ground state optimization. One simply targets a state whose energy is closest to the energy of interest in the effective Hamiltonian $\mathcal H$. 
However, this procedure does not work in general. Problems occur when there are degeneracies in the spectrum or whenever there is a high density of states. 
Then, root flipping can occur which means that the optimization oscillates between different states.
There are various methods to avoid these problems.\cite{dorando_2007,hu_2015a,baiardi_2019,doriol_2008,wodraszka_2012,zhao_2016,tran_2019a,kosloff_1986}
In the following, we describe three methods. The first two methods are based on making use of the previously computed states whereas the third method is based on state averaging.
Two of the methods will be combined in \autoref{sec:appl_ch3cn}.
We note that, as the used sweep algorithm simply is a straightforward generalization of the DMRG algorithm, any improvement from the DMRG literature\cite{dorando_2007,baiardi_2019}
can straightforwardly be implemented for TTNS.
Here, we stick to more basic methods in order to present the overall methodology.

\subsubsection{Projecting out previously computed states}
\label{sec:theory_projecting_out}

A very common approach is to project the previously computed states $\ket{I}$ out of the solution space.\cite{kosloff_1986,baiardi_2019}
For that, we first define the projection operator $\hat P$ on the previously computed states $\ket{I}$ as 
\begin{equation}
    \hat P =  \sum_I \ketbra{I}{I}.\label{eq:H_shift_P}
\end{equation}
The Hamiltonian is then modified to 
\begin{equation}
\hat{H}^P = (\hat1 - \hat P) \hat H (\hat1 - \hat P).
\end{equation}
This approach has some practical problems. For example, $\hat H^P$ includes a null space that may interfere with the state of interest. Furthermore, $\hat H^P$ is more complicated than the approach presented in the next section and numerical difficulties arise once $\{\ket I\}$ are not fully orthogonal.

\subsubsection{Shifting the spectrum}
\label{sec:theory_spectral_shift}

As an alternative to the projection method, the projector can be used not to project the states out but to shift the energies $E_I$ of the states $\ket I$ by an amount $S$:\cite{wouters_2014d}

\begin{equation}
  \hat{H}^S = \hat H + \sum_I \newi{(E_I+S)} \ketbra{I}{I}. \label{eq:H_shift}
\end{equation}
If $S$ is larger than the difference between the energy levels of the states $\ket{I}$ and the targeted state, the targeted states becomes the ground state of $\hat H^S$ and can thus easily be retrieved.

Inserting \autoref{eq:H_shift} into \autoref{eq:MF}, the matrix elements of the effective Hamiltonian $\mathcal H^S$ take the form of
\begin{align}
\mathcal H^S_{AB} &= \matrixe{\Phi^{[1]}_A}{H}{\Phi^{[1]}_B} +  \sum_I \newi{(E_I+S)} \braket{\Phi^{[1]}_A}{I} \braket{I}{\Phi^{[1]}_A} \\
&= \mathcal H_{AB}  +  \sum_I \newi{(E_I+S)}  \braket{\Phi^{[1]}_A}{I} \braket{I}{\Phi^{[1]}_A}.
\end{align}
Note that after the optimization, the newly computed eigenstate is orthogonal to the previously computed states $\{\ket I\}$. Then, $\braket{\Phi^{[1]}_A}{I}$ typically approaches zero and $\mathcal H^S$ approaches $\mathcal H$.

\subsubsection{State average calculations}
\label{sec:theory_state_av}

Another, well-known way to obtain excited states is to describe several eigenstates simultaneously with the same tensor network. This is known as state averaging.\cite{Culot1994,manthe_2008,doriol_2008,hammer_2012}
It can be achieved by adding to the root node an additional ``physical'' bond that indicates the particular state.
This has been used both for MCTDH and ML-MCTDH, and there, it is easily implemented. \cite{manthe_2008,doriol_2008,hammer_2012}

For the sweep algorithm in TTNS, the implementation is less obvious as the root node changes during the sweep.
In principle, the state dimension could stay on one particular node. However, the states then become nonorthogonal once the canonical form changes. 
We found that this leads to arbitrarily small weights of the particular states and thus an  unstable optimization.

\begin{figure}
  \includegraphics{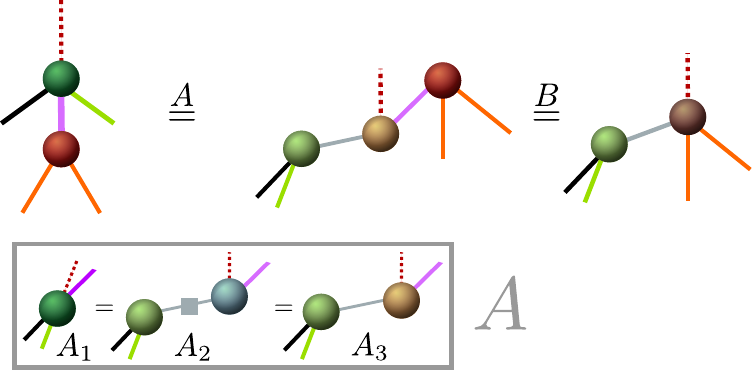}
  \caption{Canonicalization with state averaging. For clarity, the bonds are colored. The dashed bond represent the state dimension. The substeps for step $A$ are shown in the lower panel (gray rectangle). 
  The gray square in step $A_2$ corresponds to a diagonal matrix (the singular values). See text for details.}
  \label{fig:sa}
\end{figure}

To make the optimization stable, it is required to move the state dimension from one node to another. This has already been used in DMRG calculations by means of taking averages of density matrices.\cite{ronca_2017a}
Here, instead of using density matrices, we use a more simpler and less costly approach based on a SVD.
The procedure is described in \autoref{fig:sa}: In step $A$, the old root node $G$ with the state dimension is first transposed (step $A_1$; note the change of the order of the bonds in \autoref{fig:sa}) and then decomposed using a SVD (steps $A_2$ and $A_3$):
\begin{equation}
\underbrace{\matr G}_{A_1} = \underbrace{\matr U \matr s \matr V^\dagger}_{A_2} = \underbrace{\matr U\widetilde{\matr V}}_{A_3}.\label{eq:sa_svd}
\end{equation}
The remaining dimensions (the rows of $\matr U$) are now separated from the state dimension and the dimension connecting to the next node (the columns of $\widetilde{\matr V}$). 
In the final step $B$, $\widetilde{ \matr V}$ is absorbed into the neighboring node, which becomes the new root node.

One drawback of this procedure is that the transposition of the previous root tensor in step $A_1$ changes the ordering of dimensions in the tensor network. 
This means that the initial and final tensor networks shown in \autoref{fig:sa} are not related via a gauge transform. 
Therefore, the required bond dimension to represent the same physical state to a given accuracy changes (in the example in \autoref{fig:sa}, the dimension of the purple bond differs from that of the gray bond) and often grows. 
To avoid a grow of the bond dimension, it is thus required to truncate it (by means of discarding some singular values/some rows of $\widetilde{\matr V}$). 
Then, $\ket{\Psi}$ \emph{varies} whenever the canonical form changes and the energy of the state depends on the position of the root node in the tree. Typically, the energy is lowest in the center of a MPS\cite{ronca_2017a} but this is less obvious in a tree.
Furthermore, it is crucial to diagonalize the new root node once the state dimension has moved. Canonicalizing from one node to a not directly connected node without diagonalization will deteriorate the state if the bond dimension is truncated.

\subsection{Adaption of the bond dimension}
\label{sec:theory_bond_adaption}

A drawback of complicated tensor network states such as TTNS is the number of input parameters.
For example, the required bond dimensions for each node need to be set.
While this can be done in an iterative manner by observing the natural weights (eigenvalues of the density matrix) after the TTNS is converged,
this procedure is cumbersome. Instead, we here make use of an adaptive bond dimension during the optimization procedure.\cite{legeza_2003a,ml_spawning_gatti_2017,dp_wodraszka_2017,dpmctdh_tannor_2017}
For state-averaged states, this can easily be implemented by truncating the singular values $s_i$ in 
\autoref{eq:sa_svd} such that only singular values larger than some parameter $\epsilon$ are included.\footnote{In practice, the bond dimension is increased by one after every adaption. This avoids oscillations of the bond dimension.}
For TTNS without state averaging, bond adaption is achieved by observing the natural weights for the bond between the root node $\matr A$ and the neighboring node $\matr B$ that will become the new root node during the sweep. This allows to reduce the bond dimension (by discarding some natural orbitals). In the current implementation, increasing the bond dimension is achieved by performing a SVD of the matricized, combined tensor $\matr{AB}$. Improved methods may be used in future.\cite{legeza_2003a}
The two-site algorithm offers the most straightforward and robust way to adapt the bond between two nodes. It is, however, also more costly.

\subsection{Finding good trees}
\label{sec:theory_disentangling}

\begin{figure}
    \includegraphics[width=\columnwidth]{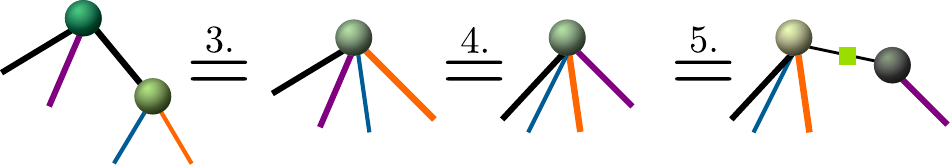}
  \caption{Procedure to optimize a tree. After randomly selecting a tensor (1.; not shown) and canonicalizing to it (2.; not shown), it is contracted with a randomly chosen neighbor (3.). Then (4.), their dimensions are permuted. Finally (5.), a SVD along a random matricization is performed to check the required bond dimension.}
  \label{fig:disentangling}
\end{figure}

Another drawback of TTNS over simpler methods is that it is not always obvious how to set up an optimal tree, meaning a tree ``topology'' that has low bond dimensions for a given accuracy.\cite{ml_spawning_gatti_2017}
The same holds for the ordering of the physical dimensions in a MPS.\cite{Chan2002,Olivares-Amaya2015}
Here, we propose a simple way to improve (to ``disentangle'') the tree, meaning that smaller bond dimensions for a given accuracy are required. Other procedures based on quantum information theory are also possible.\cite{murg_2015} 
Our procedure is based on randomly permuting dimensions of neighboring tensors in the tree. Once some eigenstate has been obtained with an initial tree that is based on an educated guess, 
the following (see also \autoref{fig:disentangling}) is iterated many times:
\begin{enumerate}
  \item Randomly select tensor $\matr X$ and a neighboring tensor $\matr Y$.
  \item Canonicalize to $\matr X$. 
  \item Contract $\matr X$ with $\matr Y$ to get the combined tensor $\matr Z$ (see \autoref{fig:disentangling}).
  \item Randomly permute the dimensions of $\matr Z$, e.g., $Z_{ijkl} \to \widetilde Z_{klji}$ (see \autoref{fig:disentangling}).\label{dis:permut}
  \item Perform a SVD for a randomly chosen matricization of $\widetilde{\matr Z}$: $\widetilde Z_{klji} = \sum_{x} U_{kljx} s_x V_{xi} =  \sum_{x} U_{kljx} \widetilde V_{xi}$ (see \autoref{fig:disentangling}).
\item Set the bond dimension of this new configuration as the number of singular values that are larger than some parameter $\epsilon$. If this is smaller than the previous bond dimension: Accept the new configuration, replacing $\matr X$ and $\matr Y$ by $\matr U$ and $\widetilde{\matr V}$. Otherwise, discard it.\label{dis:select}
\end{enumerate}
For particular tensors $\matr X$ and $\matr Y$, steps \ref{dis:permut} to \ref{dis:select} can be repeated several times (microiterations) before two new tensors are selected (macroiterations).

The optimization procedure makes use of ``greedy'' optimization. 
As new configurations are only accepted if they actually lower the bond dimension, the optimization likely gets stuck in a local minimum. This is because only neighboring tensors are combined. However, the algorithm can simply be improved by accepting worse configurations with some probability, making it similar to simulated annealing.\cite{weise_book}

Since two nodes need to be contracted for this algorithm, it scales as $n^{2d}$. This is larger than the normal scaling of the optimization, which is $n^{d+1}$. However, by computing eigenvalues of the reduced density matrix instead of computing a SVD and using iterative eigensolvers, the scaling can be reduced. At any rate, the optimization procedure is very fast because the prefactor of the scaling is orders of magnitudes smaller, compared to that of the eigenvalue optimization.

\section{Applications}
\label{sec:appl_ch3cn}

In the following, we present results for a quartic force field of acetonitrile,\cite{begue_2005,avila_smolyak_2011}
which is a twelve-dimensional problem.
This system has been established as a benchmark model, allowing for a comparison to a variety of methods.\cite{avila_smolyak_2011,Rakhuba2016a,dp_wodraszka_2017,Leclerc2014,thomas_hcp_2015,Baiardi2017}
Here, we compare both TTNS and MPS against Smolyak quadrature performed by Avila and Carrington\cite{avila_smolyak_2011} and previous tensor train/MPS calculations performed by Rakhuba and Oseledets.\cite{Rakhuba2016a}
The goal is to compute the lowest 84 eigenstates of the Hamiltonian to sub-$\unit{cm^{-1}}$ accuracy.
While the correlation in this system is not very strong, difficulties arise due to a high density of states with many nearly-degenerate states, combined with the high accuracy demand. 
We note that the desired accuracy only serves as benchmark purpose, to make the problem more difficult and to compare with previous highly accurate calculations. 
The system parameters are described in \autoref{sec:appl_ch3cn_par}. \autoref{sec:appl_ch3cn_met} describes the particular methodology and combinations of the algorithms presented in \autoref{sec:theory}. The results are given in \autoref{sec:appl_ch3cn_res}.

\subsection{System parameters}
\label{sec:appl_ch3cn_par}

\begin{figure}
     \includegraphics{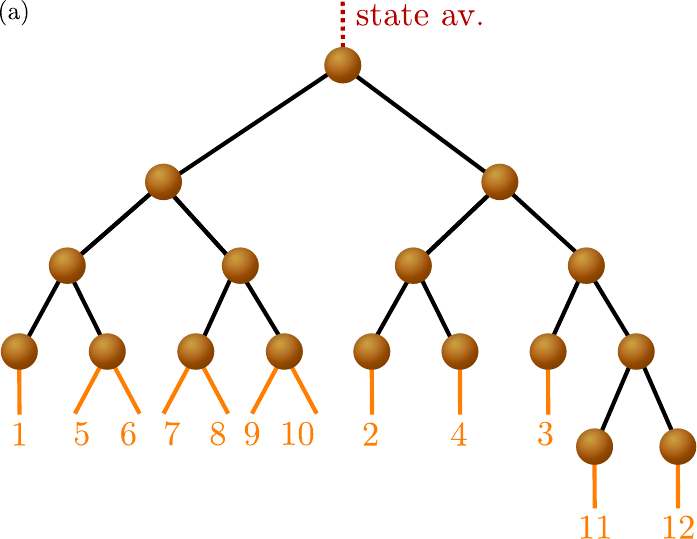}
     \includegraphics{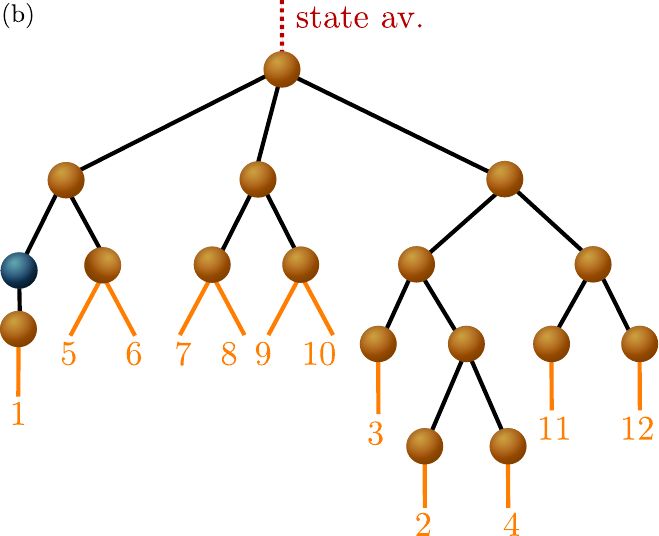}
     \caption{Used tree tensor network states employed for the acetonitrile calculations. The numbers denote the particular physical dimension. (a) is the primarily used tree and (b) the optimized tree. The blue node denotes a redundant node/tensor.}
     \label{fig:ttns_ch3cn}
\end{figure}

We took the parameters of the quartic force field from the Heidelberg MCTDH package.\cite{mctdh_package}
The $J=0$ Hamiltonian, using normal coordinates and neglecting cross terms in the kinetic energy operator, has then 323 product terms. 
We used the same Gauß-Hermite discrete variable representation (DVR)\cite{tannor_book}
employed by Leclerc and Carrington and Rakhuba and Oseledets.\cite{Leclerc2014,Rakhuba2016a}
The basis sizes were $N_i\in \{9, 7, 9, 9, 9, 9, 7, 7, 9, 9, 27, 27\}$ for the modes with harmonic frequencies (in $\unit{cm^{-1}}$) $\omega_1=3065$, $\omega_2=2297$, $\omega_3=1413$, $\omega_4=920$, $\omega_5=\omega_6=3149$, $\omega_7=\omega_8=1487$, $\omega_9=\omega_{10}=1061$, $\omega_{11}=\omega_{12}=361$.
The TTNS we used is shown in \autoref{fig:ttns_ch3cn} in panel (a). The particular way of setting up the tree roughly followed previous investigations with similar methods.\cite{thomas_hcp_2015,dp_wodraszka_2017}
For the MPS, we used the same basis ordering as used by Rakhuba and Oseledets,\cite{Rakhuba2016a} namely, we ordered the modes according to their frequency. Similar to the findings of Rakhuba and Oseledets, we could not find particularly improved results for different orderings.

\subsection{Methodology}
\label{sec:appl_ch3cn_met}

To obtain the lowest 84 eigenstates, we iterated the following procedure, both for TTNS and MPS:
\begin{enumerate}
  \item Do a (loose) state average calculation with 20 states for maximal seven sweeps (as shown in \autoref{fig:sweep_schedule}). This gives a rough estimate of the eigenstates.
  \item Cluster the 20 states in groups whose energy differ by maximal $\unit[2]{cm^{-1}}$. Discard the last cluster as this may overlap with higher lying states.
  \item For each cluster, do an additional (refined) state average calculation for maximal ten sweeps. This gives the converged states.
\end{enumerate}
The convergence criterion for the sweep was a relative difference of $10^{-9}$ in the energy, compared to the energy of the previous sweep.
Throughout, the Hamiltonian was shifted by the previously computed states (see \autoref{eq:H_shift}). We used a shift of $S=\unit[5000]{cm^{-1}}$.
In each calculation, we adapted the bond dimension every fourth sweep (see \autoref{sec:theory_bond_adaption}). Besides using a singular value threshold $\epsilon$ for the bond adaption, we fixed the maximally allowed bond dimension $n_\text{max}$ and also set a minimal bond dimension of $n_\text{min}=3$.
We performed calculations with a bond dimension threshold of either $\epsilon = 10^{-4}$ or $10^{-5}$ and a maximal bond dimension of either $n_\text{max}=25$, $40$ or $50$.
For the state average calculations, the accuracy  of the eigenstates depends on which node they are evaluated. Hence, after convergence of the loose state average calculation (with 20 states), we selected the root node that gave the lowest energies.

In principle, one could also avoid the state average calculations at all and perform optimizations state by state. However, state averaging accelerates the optimization procedure as fewer optimizations are necessary, even though larger bond dimensions are required in state average calculations.
Another way to compute the spectrum would be to perform just one big state average calculation. This has been done for \ce{CH3CN} for computing the lowest 69 states with ML-MCTDH.\cite{dp_wodraszka_2017} However, this approach requires very big bond dimensions as all eigenstates need to be described by the same tree.
The more eigenstates needed to be computed, the tighter the demands on the bond dimension.

As proof of concept, after having obtained eigenstates, we optimized the tree as explained in \autoref{sec:theory_disentangling}. For that, we arbitrarily chose the 10th excited state and optimized the tree using the ``greedy'' algorithm (only accepting new tensor configurations if the bond dimension is reduced). We did an overly exhaustive optimization and performed 4000 macroiterations (randomly selecting a tensor and a neighbor) and within each macroiteration we performed 190 microiterations (randomly permuting dimensions).

\newi{We also compared to ML-MCTDH-based optimizations, both for the ground state and for a state-averaged computation minimizing the lowest 13 states. 
For that, we used a fixed number of single-particle functions in each layer of $n^{1,\kappa} = \{20, 20\}$,
$n^{2,\kappa} = \{8,20,6,25\}$, $n^{3,\kappa} =  \{4,6,10,8,5,6,6,10\}$ and $n^{4,\kappa} = \{6,6\}$.
The ML-MCTDH optimizations used a regularization parameter of the equations of motions of $\epsilon_r = 10^{-10}$. In improved relaxation, after each diagonalization of the root node, the SPFs were propagated until the norm of their time-derivative was below $10^{-9}$. The accuracy of the propagators were set to $10^{-9}$ as well.
}

The program we used is implemented in Python and makes use of NumPy\cite{numpy} and SciPy.\cite{scipy}
The routines for computing a matrix-vector product with matrices decomposed in SOP form are implemented in C++ and are taken from Ref.~\citenum{pW_tannor_2016} (see appendix therein). 
As diagonalizer we use Davidson's method,\cite{davidson_1975}
as implemented in PySCF.\cite{pyscf}

\subsection{Results and Discussion}
\label{sec:appl_ch3cn_res}

In the following, we present the results and compare them to results based on Smolyak quadrature,  performed by Avila and Carrington,\cite{avila_smolyak_2011}
and to results based on tensor trains (TT), performed by Rakhuba and Oseledets.\cite{Rakhuba2016a}

Rakhuba and Oseledets obtained a zero point vibrational energy (ZPVE) of $\unit[9837.4063]{cm^{-1}}$ and Avila and Carrington a ZPVE of $\unit[9837.4073]{cm^{-1}}$ (as given by Rakhuba and Oseledets). 
Despite using the same basis, we were not able to reproduce the ZPVE from Rakhuba and Oseledets. 
The best (i.e. smallest) value we could obtain with our code is $\unit[9837.4069]{cm^{-1}}$.
We confirmed this value with an independent standard MCTDH calculation (using the Heidelberg package\cite{mctdh_package})
and an exhaustive number of single particle functions (all natural weights were less than $10^{-11}$). We speculate that the (very minor) discrepancies between our results and those from Rakhuba and Oseledets exist because they fitted the potential to a tensor train (matrix product operator) format.
For this reason, in order to get very accurate reference values for the spectrum to compare to,
we performed an additional calculation without the approximation of a fit. For that, we used an MPS with bond dimension of $n_\text{max}=100$ without bond adaption and without state averaging. 

\begin{figure*}
\includegraphics{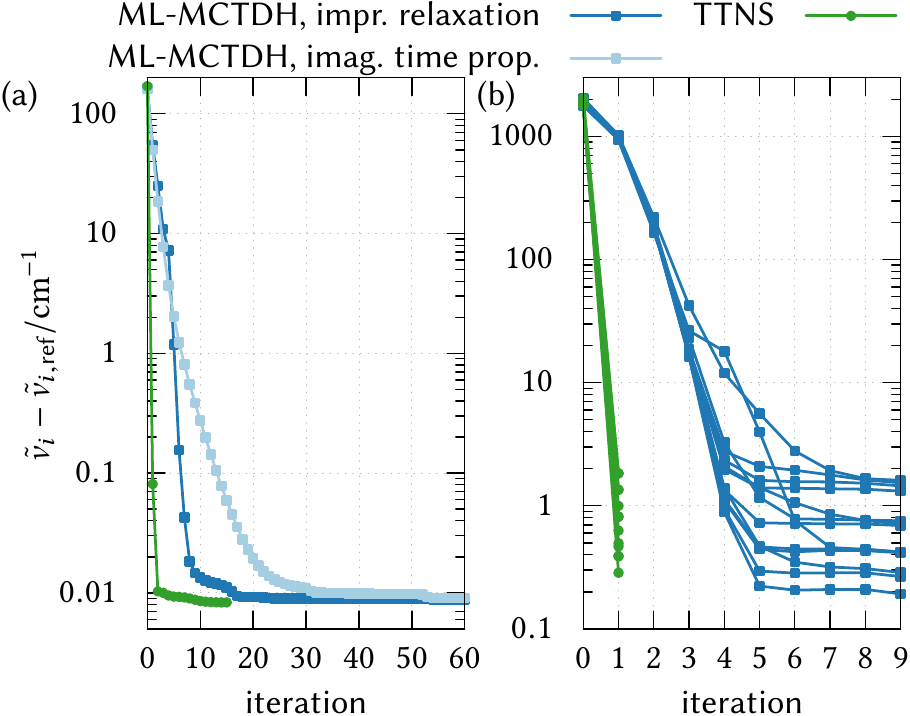}
     \caption{Absolute errors of the ground state (left panel) and the lowest 13 states (state averaged, right panel) of acetonitrile, comparing optimization based on
     tree tensor network state (TTNS) optimization (green, showing the error after each sweep), the multilayer multi-configuration time-dependent Hartree method (ML-MCTDH) with improved relaxation (dark blue, showing the error after each diagonalization) and  ML-MCTH with imaginary time propagation (pale blue, showing the error after each time step of $\unit[1]{fs}$).
     All calculations used the same tensor sizes (see text). Thus, the ground state calculation  is more accurate.
     }\label{fig:CH3CN_ML_vs_TTNS}
\end{figure*}

\newi{The performance of the TTNS-based optimization, compared to that of ML-MCTDH is shown in \autoref{fig:CH3CN_ML_vs_TTNS}, both for the ground state and the lowest 13 states (state averaged). Already after three TTNS iterations is the ground state converged to about $\unit[0.01]{cm^{-1}}$. In contrast, ML-MCTDH improved relaxation requires 17 iterations and an imaginary time propagation $\unit[34]{fs}$ 
in order to reach the same accuracy. 
State-average optimizations show a similar behavior. 
There, the TTNS optimization is already converged after just one iteration whereas the ML-MCTDH optimization requires 8 iterations to reach a similar accuracy ($\unit[\sim 1]{cm^{-1}}$, using the same tensor sizes as for the ground state minimization).
However, due to the approximate canonicalization for state-averaged root nodes (see \autoref{sec:theory_state_av}) in TTNS, the ML-MCTDH calculations can give slightly higher accuracies for a given bond dimension. Nevertheless, the state-average TTNS calculation can always be followed by additional state-specific calculations using, e.g., shifted Hamiltonians (see \autoref{sec:theory_spectral_shift}). 
}

\newi{
It would be useful to compare runtimes instead of iterations. However, our ML-MCTDH code is not fully optimized and the performance of the propagation in ML-MCTDH is very sensitive the choice of propagator. Nonetheless, one TTNS iteration is approximately comparable to one ML-MCTDH iteration with respect to runtime\footnote{For TTNS, each iteration requires the diagonalization of each tensor and local (from one node to a neighboring node) transformation of the Hamiltonian whereas for ML-MCTDH with improved relaxation, each iteration requires imaginary time propagation of the single particle functions, followed by a global transformations of the Hamiltonian (creation of the mean-fields), computation and inversion of density matrices and diagonalization of the root node.}
and the TTNS optimization converges much faster such that it should also be faster in runtime.
}

\newi{We now discuss the accurate state-average TTNS optimizations for the whole spectrum. 
Also in this case,} we noticed a very rapid convergence of the minimization. \newi{Similar to the previous observations, f}or the loose calculations with 20 averaged states, starting from a random initial state, already after just one sweep were the first 14 eigenstates converged to within a few $\unit{cm^{-1}}$.
Thus, no two-site algorithm is necessary in this case. 

\begin{figure}
     \includegraphics{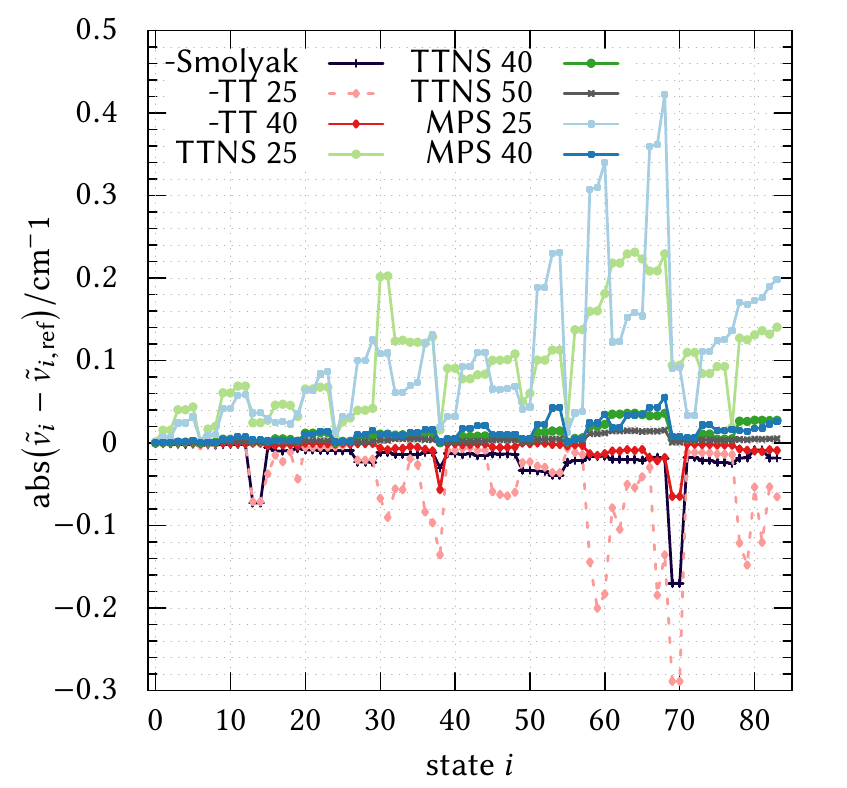}
     \caption{Absolute errors of the energy levels for Smolyak,\cite{avila_smolyak_2011}
       tensor train (TT),\cite{Rakhuba2016a}
     tree tensor network state (TTNS) and matrix product state (MPS) calculations.
     The number behind the names represent the maximally allowed bond dimension. See \autoref{tab:CH3CN_levels} for more details. For clarity, the Smolyak and TT results are plotted on the negative  part of the ordinate, even though absolute errors are shown.}
     \label{fig:CH3CN_errors}
\end{figure}

The energy levels and errors are shown in Table \ref{tab:CH3CN_levels}. Note that negative errors for the TT calculations do not mean that the results are more accurate (because a fit of the potential is used; see above). The absolute errors are also plotted in \autoref{fig:CH3CN_errors}.

The obtained accuracy is very good. For the TTNS with $n_\text{max}=50$, the maximal absolute error in the levels is $\unit[0.015]{cm^{-1}}$. For the TTNS with $n_\text{max}=40$, the maximal error is 
 $\unit[0.036]{cm^{-1}}$. The errors of the MPS are comparable, although slightly larger for $n_\text{max}=25$, compared to the same TTNS.  The TT calculations from Rakhuba and Oseledets with fixed bond dimension give very similar errors, compared to our MPS calculations with variable bond dimension.

\begingroup
\LTcapwidth=.85\textwidth \begin{longtable*} {lddddddddd}
  \caption{Vibrational levels $\tilde \nu_i$ and absolute errors  (in $\unit{cm^{-1}}$) of acetonitrile.
  The error is represented with respect to calculations from a matrix product state (MPS) with (fixed) bond dimension of $n=100$. The tree tensor network states (TTNS) and MPS calculations were performed with an adaptive bond dimension with parameter $\epsilon$ (\autoref{sec:theory_bond_adaption}) and  maximal bond dimension of $n_\text{max}$.
  The results from the Smolyak quadrature are taken from Ref.~\citenum{avila_smolyak_2011},
  The results from the tensor trains (TT) with fixed bond dimension are taken from Ref.~\citenum{Rakhuba2016a}.
  The assignment is according to Ref.~\citenum{avila_smolyak_2011}.
  \label{tab:CH3CN_levels}
}\\
  \toprule
 	&	\multicolumn{1}{c}{Ref. MPS}	&	\multicolumn{1}{c}{Smolyak}	&	\multicolumn{1}{c}{TT}	&		\multicolumn{1}{c}{TT}&		\multicolumn{1}{c}{TTNS}&		\multicolumn{1}{c}{TTNS}&		\multicolumn{1}{c}{TTNS}	&		\multicolumn{1}{c}{MPS}&		\multicolumn{1}{c}{MPS}\\ 
\multicolumn{1}{r}{$n_\text{max}=$}    	&	     \multicolumn{1}{c}{$100$}   &               	&	\multicolumn{1}{c}{$40$}	&	\multicolumn{1}{c}{$25$}	&		\multicolumn{1}{c}{$25$}&		\multicolumn{1}{c}{$40$} &		\multicolumn{1}{c}{$50$}	&	\multicolumn{1}{c}{$25$}&		\multicolumn{1}{c}{$40$} \\
\multicolumn{1}{r}{$\epsilon=$}     	&	       &               	&	&  &		\multicolumn{1}{r}{$10^{-4}$}&		\multicolumn{1}{r}{$10^{-5}$} &		\multicolumn{1}{r}{$10^{-5}$}	&	\multicolumn{1}{r}{$10^{-4}$}&	\multicolumn{1}{r}{$10^{-5}$} \\ \cline{3-10}
Level $i$&	\multicolumn{1}{c}{$\tilde \nu_{i,\text{ref}}/\unit{cm^{-1}}$} & \multicolumn{7}{c}{$(\tilde \nu_i - \tilde \nu_{i,\text{ref}})/ \unit{cm^{-1}}$}\\
\midrule
                ZPVE	&	9837.4069	&	0.0004	&	-0.0006	&	0.0004	&	0.0013	&	0.0001	&	0.0000	&	0.0015	&	0.0001 \\ 
          $\nu_{11}$	&	360.990	&	0.001	&	0.000	&	0.000	&	0.015	&	0.001	&	0.000	&	0.007	&	0.001 \\ 
	&	360.990	&	0.001	&	0.000	&	0.000	&	0.016	&	0.001	&	0.000	&	0.007	&	0.001 \\ 
         $2\nu_{11}$	&	723.179	&	0.002	&	0.001	&	0.000	&	0.040	&	0.001	&	0.001	&	0.024	&	0.002 \\ 
	&	723.179	&	0.002	&	0.001	&	0.000	&	0.040	&	0.001	&	0.001	&	0.024	&	0.002 \\ 
         $2\nu_{11}$	&	723.825	&	0.002	&	0.001	&	0.001	&	0.044	&	0.001	&	0.001	&	0.031	&	0.003 \\ 
           $\nu_{4}$	&	900.659	&	0.003	&	-0.001	&	-0.003	&	0.001	&	0.000	&	0.000	&	0.003	&	0.001 \\ 
           $\nu_{9}$	&	1034.124	&	0.002	&	0.000	&	0.001	&	0.017	&	0.001	&	0.000	&	0.009	&	0.001 \\ 
	&	1034.124	&	0.002	&	0.000	&	0.001	&	0.020	&	0.001	&	0.000	&	0.010	&	0.001 \\ 
         $3\nu_{11}$	&	1086.552	&	0.002	&	0.000	&	0.000	&	0.061	&	0.005	&	0.002	&	0.042	&	0.004 \\ 
         $3\nu_{11}$	&	1086.552	&	0.002	&	0.001	&	0.001	&	0.061	&	0.004	&	0.002	&	0.041	&	0.004 \\ 
         $3\nu_{11}$	&	1087.774	&	0.002	&	0.001	&	0.001	&	0.069	&	0.005	&	0.002	&	0.058	&	0.008 \\ 
	&	1087.774	&	0.002	&	0.001	&	0.001	&	0.069	&	0.005	&	0.002	&	0.058	&	0.007 \\ 
  $\nu_{4}+\nu_{11}$	&	1259.809	&	0.073	&	0.000	&	-0.071	&	0.024	&	0.001	&	0.001	&	0.036	&	0.004 \\ 
	&	1259.809	&	0.073	&	0.000	&	-0.071	&	0.024	&	0.001	&	0.001	&	0.036	&	0.004 \\ 
           $\nu_{3}$	&	1388.967	&	0.006	&	0.004	&	0.038	&	0.027	&	0.002	&	0.000	&	0.029	&	0.002 \\ 
  $\nu_{9}+\nu_{11}$	&	1394.679	&	0.010	&	0.003	&	0.015	&	0.046	&	0.005	&	0.002	&	0.025	&	0.003 \\ 
	&	1394.679	&	0.010	&	0.003	&	0.023	&	0.047	&	0.005	&	0.002	&	0.026	&	0.003 \\ 
  $\nu_{9}+\nu_{11}$	&	1394.898	&	0.009	&	0.002	&	0.011	&	0.046	&	0.004	&	0.001	&	0.023	&	0.003 \\ 
  $\nu_{9}+\nu_{11}$	&	1397.680	&	0.007	&	0.004	&	0.044	&	0.032	&	0.002	&	0.001	&	0.035	&	0.003 \\ 
         $4\nu_{11}$	&	1451.093	&	0.008	&	0.000	&	-0.006	&	0.065	&	0.012	&	0.002	&	0.063	&	0.010 \\ 
	&	1451.093	&	0.008	&	0.000	&	-0.006	&	0.065	&	0.012	&	0.002	&	0.063	&	0.010 \\ 
         $4\nu_{11}$	&	1452.818	&	0.009	&	0.001	&	-0.005	&	0.068	&	0.012	&	0.003	&	0.084	&	0.014 \\ 
	&	1452.818	&	0.009	&	0.001	&	-0.005	&	0.068	&	0.012	&	0.003	&	0.086	&	0.014 \\ 
         $4\nu_{11}$	&	1453.394	&	0.009	&	0.001	&	-0.005	&	0.006	&	0.000	&	0.000	&	0.012	&	0.002 \\ 
           $\nu_{7}$	&	1483.219	&	0.010	&	0.001	&	-0.003	&	0.026	&	0.002	&	0.001	&	0.032	&	0.002 \\ 
	&	1483.220	&	0.009	&	0.001	&	-0.002	&	0.030	&	0.002	&	0.001	&	0.032	&	0.002 \\ 
 $\nu_{4}+2\nu_{11}$	&	1620.199	&	0.023	&	-0.001	&	-0.021	&	0.040	&	0.008	&	0.001	&	0.100	&	0.010 \\ 
	&	1620.199	&	0.023	&	-0.001	&	-0.021	&	0.040	&	0.008	&	0.001	&	0.100	&	0.010 \\ 
 $\nu_{4}+2\nu_{11}$	&	1620.744	&	0.023	&	-0.001	&	-0.020	&	0.042	&	0.008	&	0.002	&	0.125	&	0.015 \\ 
  $\nu_{3}+\nu_{11}$	&	1749.519	&	0.011	&	0.006	&	0.067	&	0.201	&	0.011	&	0.003	&	0.108	&	0.010 \\ 
	&	1749.519	&	0.011	&	0.008	&	0.090	&	0.202	&	0.011	&	0.003	&	0.109	&	0.011 \\ 
 $\nu_{9}+2\nu_{11}$	&	1756.412	&	0.014	&	0.007	&	0.056	&	0.123	&	0.009	&	0.005	&	0.061	&	0.010 \\ 
 $\nu_{9}+2\nu_{11}$	&	1756.412	&	0.014	&	0.007	&	0.057	&	0.124	&	0.010	&	0.005	&	0.061	&	0.010 \\ 
 $\nu_{9}+2\nu_{11}$	&	1757.119	&	0.014	&	0.004	&	0.019	&	0.122	&	0.009	&	0.005	&	0.070	&	0.012 \\ 
	&	1757.119	&	0.014	&	0.005	&	0.026	&	0.122	&	0.009	&	0.005	&	0.073	&	0.012 \\ 
 $\nu_{9}+2\nu_{11}$	&	1759.760	&	0.012	&	0.008	&	0.084	&	0.121	&	0.012	&	0.004	&	0.122	&	0.016 \\ 
	&	1759.760	&	0.012	&	0.010	&	0.097	&	0.129	&	0.013	&	0.004	&	0.131	&	0.017 \\ 
          $2\nu_{4}$	&	1785.177	&	0.030	&	-0.057	&	-0.136	&	0.016	&	0.000	&	0.000	&	0.019	&	0.001 \\ 
         $5\nu_{11}$	&	1816.786	&	0.013	&	0.001	&	-0.009	&	0.090	&	0.004	&	0.001	&	0.032	&	0.005 \\ 
	&	1816.786	&	0.013	&	0.001	&	-0.009	&	0.090	&	0.004	&	0.001	&	0.032	&	0.005 \\ 
         $5\nu_{11}$	&	1818.938	&	0.014	&	0.002	&	-0.007	&	0.078	&	0.008	&	0.002	&	0.092	&	0.018 \\ 
         $5\nu_{11}$	&	1818.939	&	0.013	&	0.002	&	-0.006	&	0.078	&	0.008	&	0.002	&	0.092	&	0.018 \\ 
         $5\nu_{11}$	&	1820.016	&	0.015	&	0.001	&	-0.009	&	0.083	&	0.008	&	0.002	&	0.109	&	0.021 \\ 
	&	1820.016	&	0.015	&	0.001	&	-0.009	&	0.083	&	0.009	&	0.002	&	0.110	&	0.021 \\ 
  $\nu_{7}+\nu_{11}$	&	1844.245	&	0.013	&	0.005	&	0.059	&	0.100	&	0.007	&	0.003	&	0.065	&	0.010 \\ 
  $\nu_{7}+\nu_{11}$	&	1844.316	&	0.014	&	0.006	&	0.063	&	0.100	&	0.007	&	0.004	&	0.065	&	0.010 \\ 
	&	1844.317	&	0.013	&	0.005	&	0.064	&	0.101	&	0.007	&	0.004	&	0.066	&	0.010 \\ 
  $\nu_{7}+\nu_{11}$	&	1844.676	&	0.014	&	0.005	&	0.060	&	0.108	&	0.007	&	0.004	&	0.068	&	0.011 \\ 
   $\nu_{4}+\nu_{9}$	&	1931.514	&	0.033	&	0.000	&	-0.024	&	0.050	&	0.004	&	0.001	&	0.041	&	0.005 \\ 
	&	1931.514	&	0.033	&	0.001	&	-0.023	&	0.060	&	0.004	&	0.001	&	0.043	&	0.006 \\ 
 $\nu_{4}+3\nu_{11}$	&	1981.815	&	0.034	&	0.000	&	-0.028	&	0.100	&	0.013	&	0.004	&	0.188	&	0.023 \\ 
 $\nu_{4}+3\nu_{11}$	&	1981.815	&	0.035	&	0.000	&	-0.029	&	0.100	&	0.013	&	0.004	&	0.188	&	0.023 \\ 
 $\nu_{4}+3\nu_{11}$	&	1982.818	&	0.039	&	-0.002	&	-0.036	&	0.113	&	0.014	&	0.005	&	0.230	&	0.042 \\ 
	&	1982.818	&	0.039	&	-0.002	&	-0.036	&	0.113	&	0.015	&	0.005	&	0.230	&	0.043 \\ 
          $2\nu_{9}$	&	2057.044	&	0.024	&	0.004	&	0.007	&	0.026	&	0.001	&	0.001	&	0.013	&	0.002 \\ 
          $2\nu_{9}$	&	2065.265	&	0.021	&	0.002	&	-0.012	&	0.137	&	0.005	&	0.003	&	0.036	&	0.004 \\ 
	&	2065.265	&	0.021	&	0.003	&	0.014	&	0.137	&	0.006	&	0.003	&	0.038	&	0.005 \\ 
 $\nu_{3}+2\nu_{11}$	&	2111.364	&	0.016	&	0.013	&	0.144	&	0.160	&	0.020	&	0.011	&	0.307	&	0.024 \\ 
	&	2111.364	&	0.016	&	0.015	&	0.200	&	0.160	&	0.020	&	0.011	&	0.310	&	0.024 \\ 
 $\nu_{3}+2\nu_{11}$	&	2112.281	&	0.016	&	0.013	&	0.183	&	0.181	&	0.022	&	0.012	&	0.340	&	0.035 \\ 
 $\nu_{9}+3\nu_{11}$	&	2119.307	&	0.020	&	0.010	&	0.079	&	0.218	&	0.035	&	0.014	&	0.122	&	0.019 \\ 
	&	2119.307	&	0.020	&	0.010	&	0.105	&	0.218	&	0.035	&	0.014	&	0.122	&	0.019 \\ 
 $\nu_{9}+3\nu_{11}$	&	2120.521	&	0.020	&	0.008	&	0.050	&	0.229	&	0.036	&	0.015	&	0.152	&	0.033 \\ 
	&	2120.521	&	0.020	&	0.009	&	0.054	&	0.231	&	0.036	&	0.015	&	0.159	&	0.034 \\ 
 $\nu_{9}+3\nu_{11}$	&	2120.889	&	0.021	&	0.008	&	0.041	&	0.223	&	0.034	&	0.014	&	0.154	&	0.034 \\ 
 $\nu_{9}+3\nu_{11}$	&	2122.816	&	0.018	&	0.019	&	0.030	&	0.208	&	0.033	&	0.014	&	0.359	&	0.043 \\ 
	&	2122.816	&	0.018	&	0.022	&	0.185	&	0.209	&	0.033	&	0.014	&	0.362	&	0.043 \\ 
 $\nu_{9}+3\nu_{11}$	&	2123.282	&	0.019	&	0.018	&	0.136	&	0.229	&	0.036	&	0.015	&	0.423	&	0.055 \\ 
 $2\nu_{4}+\nu_{11}$	&	2142.444	&	0.170	&	-0.065	&	-0.289	&	0.094	&	0.004	&	0.001	&	0.090	&	0.008 \\ 
	&	2142.444	&	0.170	&	-0.065	&	-0.289	&	0.094	&	0.004	&	0.001	&	0.092	&	0.008 \\ 
         $6\nu_{11}$	&	2183.617	&	0.018	&	0.002	&	-0.011	&	0.110	&	0.005	&	0.004	&	0.034	&	0.006 \\ 
	&	2183.617	&	0.018	&	0.002	&	-0.011	&	0.110	&	0.005	&	0.004	&	0.034	&	0.006 \\ 
         $6\nu_{11}$	&	2186.117	&	0.021	&	0.002	&	-0.012	&	0.084	&	0.011	&	0.004	&	0.111	&	0.022 \\ 
	&	2186.117	&	0.021	&	0.002	&	-0.012	&	0.084	&	0.011	&	0.004	&	0.111	&	0.022 \\ 
         $6\nu_{11}$	&	2187.618	&	0.024	&	0.003	&	-0.013	&	0.093	&	0.005	&	0.003	&	0.125	&	0.015 \\ 
	&	2187.618	&	0.024	&	0.003	&	-0.013	&	0.093	&	0.005	&	0.003	&	0.125	&	0.015 \\ 
         $6\nu_{11}$	&	2188.119	&	0.025	&	0.003	&	-0.014	&	0.019	&	0.005	&	0.004	&	0.136	&	0.016 \\ 
 $\nu_{7}+2\nu_{11}$	&	2206.608	&	0.018	&	0.007	&	0.121	&	0.127	&	0.026	&	0.004	&	0.171	&	0.015 \\ 
 $\nu_{7}+2\nu_{11}$	&	2206.615	&	0.018	&	0.009	&	0.148	&	0.125	&	0.026	&	0.004	&	0.168	&	0.014 \\ 
 $\nu_{7}+2\nu_{11}$	&	2206.757	&	0.009	&	0.010	&	0.054	&	0.131	&	0.027	&	0.005	&	0.173	&	0.018 \\ 
	&	2206.758	&	0.008	&	0.010	&	0.120	&	0.136	&	0.027	&	0.005	&	0.176	&	0.018 \\ 
 $\nu_{7}+2\nu_{11}$	&	2207.541	&	0.018	&	0.008	&	0.053	&	0.132	&	0.027	&	0.005	&	0.190	&	0.023 \\ 
	&	2207.541	&	0.018	&	0.009	&	0.065	&	0.140	&	0.027	&	0.005	&	0.199	&	0.026 \\ 
  \bottomrule
\end{longtable*}
\endgroup

\begin{figure}
     \includegraphics{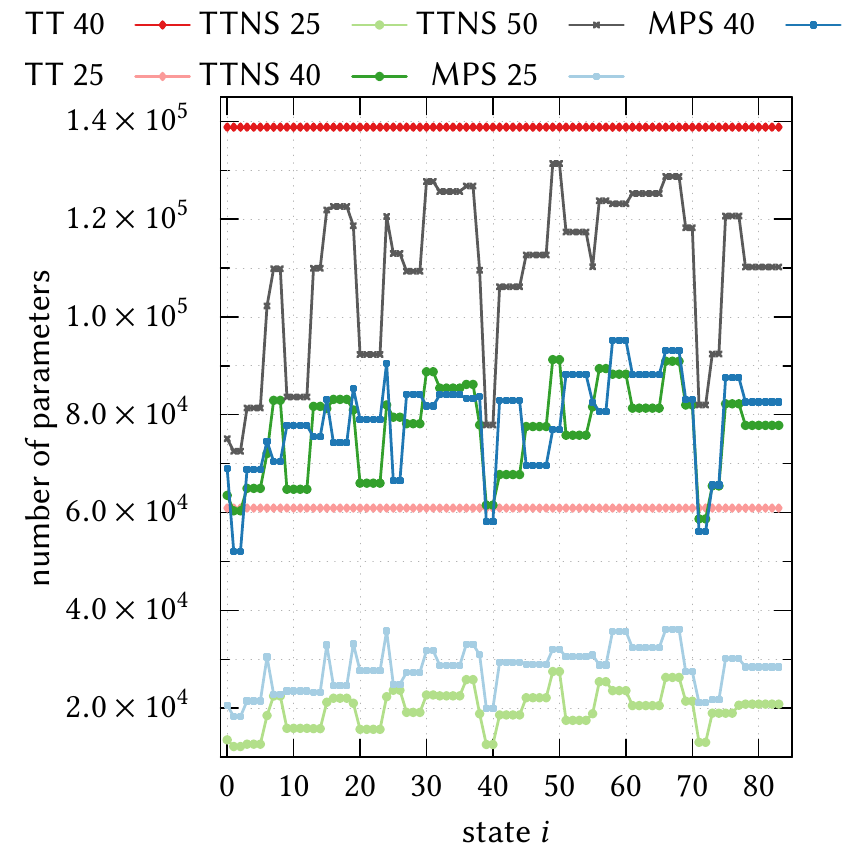}
     \caption{Total number of parameters for each state for the tensor train (TT),\cite{Rakhuba2016a}
     tree tensor network state (TTNS) and matrix product state (MPS) calculations. The number behind the names represent the maximally allowed bond dimension. The TT calculations use constant bond dimensions.}
     \label{fig:CH3CN_params}
\end{figure}

\autoref{fig:CH3CN_params} shows the number of parameters for each eigenstate.
This quantity roughly gives an estimate on the required effort of the calculations as it depends on the bond dimensions.
The adaptive bond dimension gives significant improvement, compared to a fixed bond dimension.
For $n_\text{max}=40$, the average number of parameters is around $8\cdot 10^{4}$. A fixed bond dimension of $n=40$ would give around $1.4\cdot 10^5$ number of parameters whereas a fixed bond dimension of $n=20$ would give $6\cdot 10^4$ number of parameters. Thus, an adaptive calculation with $n_\text{max}=40$ requires only marginally more parameters than a calculation with fixed $n=25$.

We note that it is not necessarily the case that the required bond dimension grows with the energy of the eigenstate. Instead, the number of parameters is roughly the same for all computed eigenstates and only oscillates significantly for some states (often to a lower number of parameters).

We now discuss the results of the proof-of-concept optimization of the tree. We note that, even though we did an overly exhaustive optimization with about $4000\times 190$ SVD calculations,
the optimization was extremely fast and only took about one minute on a standard computer. 

The optimized tree is shown in panel (b) in \autoref{fig:ttns_ch3cn}. Compared to the non-optimized tree (panel (a)), the physical dimension $3$ is now closer in the tree to dimensions $2$ and $4$. Instead of dimensions $11$ and $12$, the nodes with physical dimensions $2$ and $4$ are now in the deepest layer. Note the blue tensor in \autoref{fig:ttns_ch3cn} above dimension $1$. This tensor is redundant an can be removed from the tree.\footnote{Tensor removal and addition is not possible with the current optimization procedure but could be performed manually.}
Otherwise, there are no changes. This simply means that the initially used tree already is very good. More exhaustive global optimization procedures may result in different trees.
We point out that using an MPS as initial tree (not shown) also leads to similar ``clusters'' of tensors, including mode combination, that is, several physical dimensions on one tensor. However, the optimized trees that are based on an MPS still consists of larger linear chains and only some tensors cluster in branches of this chain. 
At any rate, the tree optimization procedure also improves very bad initial trees.

Throughout, including tree optimizations for other states and based on other initial trees, we found that three-dimensional tensors in the tree are almost exclusively selected. This empirically confirms the fact that three-dimensional tensors have the lowest scaling of number of entries with respect to dimensionality and are still possible to use within TTNS and MPS.\cite{nakatani_2013,murg_2010,gunst_2018}

\begin{figure}
     \includegraphics{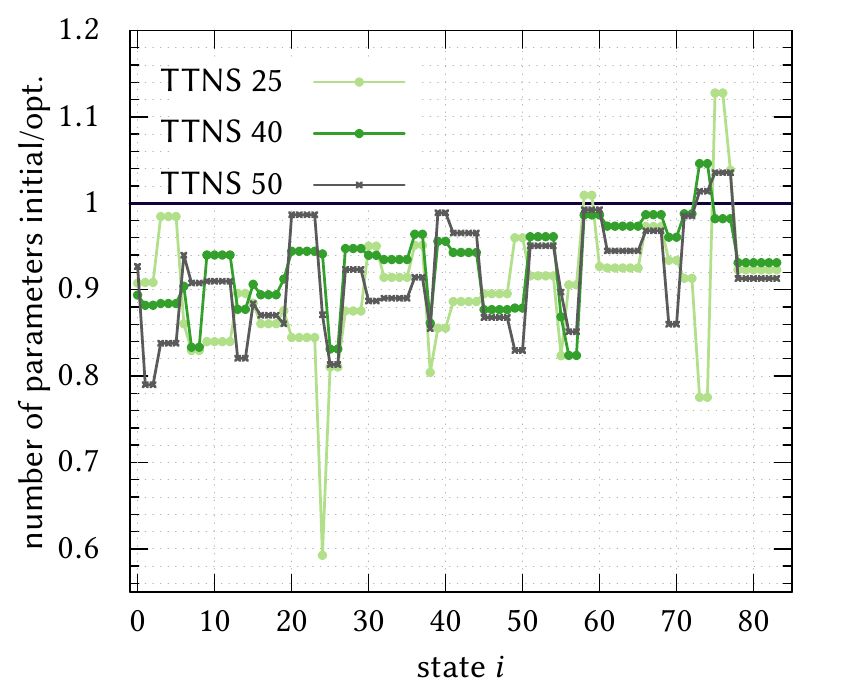}
     \caption{Ratios of number of parameters for the initial tree (see \autoref{fig:ttns_ch3cn} (a)), compared to the optimized tree (\autoref{fig:ttns_ch3cn} (b)). The number behind the names represent the maximally allowed bond dimension; compare with the labeling in ~\autoref{tab:CH3CN_levels}.}
     \label{fig:CH3CN_params_opt}
\end{figure}

Compared to the initial tree, the optimized tree results in a similar, slightly lower accuracy for the eigenstates. For example, for the setup with maximal bond dimension of $25$, the maximal absolute error of the initial tree is $\unit[0.23]{cm^{-1}}$ whereas it is $\unit[0.21]{cm^{-1}}$ in the optimized tree. The ratios of number of parameters for the initial tree, compared to the optimized tree, are shown in \autoref{fig:CH3CN_params_opt}. Even though the tree optimization is based on only the 10th eigenstate, almost all eigenstates slightly reduce the total number of parameters, which is proportional to the bond dimension, by around $10$ to $\unit[5]\%$. Only for the higher lying eigenstates is the ratio worse. 
While this result is not overly impressing, we again point out that the initial tree is already very good and that this proof-of-concept optimization can be improved by using global optimization procedures.

Interestingly, compared to MPS, TTNS need a similar amount of parameters. Only for $n_\text{max}=25$, the number of parameters of the TTNS is slightly lower and the accuracy slightly better, compared to MPS. For  $n_\text{max}=25$, the optimized tree requires around $\unit[70]\%$ of the number of parameters of the MPS. 
However for higher accuracies (larger maximal bond dimensions), it seems that, for this test case, MPS perform similarly or equally well than TTNS. 
\newi{For other systems, in particular for much higher-dimensional systems, TTNS may show a more significant advantage over MPS.}

\section{Conclusions and outlook}
\label{sec:conclusion}

In this work, we have transferred the main algorithm and the diagrammatic language
from the context of the density matrix renormalization group (DMRG) to tree tensor network states (TTNS) for computing vibrational eigenstates. This complements the well-established methodology of 
the multilayer multi-configuration time-dependent Hartree (ML-MCTDH) approach.
In particular, we showed the advantages of using the sweep algorithm from the DMRG for vibrational calculations. 

We tested TTNS and matrix product states (MPS, the \emph{ansatz} behind the DMRG)
for the 12-dimensional benchmark system acetonitrile (\ce{CH3CN}). We could obtain 
very accurate energy levels with absolute errors less than $\unit[0.04]{cm^{-1}}$ for
moderate tensor sizes (bond dimensions). 
\newi{The TTNS calculations converge much faster than ML-MCTDH-based calculations. Further, w}e showed that using adaptive tensor sizes leads
to a significant reduction of the number of parameters without jeopardizing accuracy.

Using the same code and algorithms both for MPS and TTNS allow for a direct comparison between these two tensor networks. For \ce{CH3CN}, TTNS give slightly improved accuracies with lower numbers of parameters than MPS. Compared to MPS, an optimized tree reduces the number of parameters to about $\unit[70]\%$ for medium accuracies. However, for higher accuracies,
the difference in the number of parameters is not significant and MPS perform very well and are (slightly) simpler to implement.
\newi{Nevertheless, TTNS with only three-dimensional tensors have the same computational scaling as MPS for a Hamiltonian in sum-of-product form such that it seems unlikely that TTNS will show any disadvantages, compared to MPS, besides a slightly more complicated implementation. Thus, a TTNS could mostly be favored over an MPS. In any case, more}
tests with \newi{other} systems (including more complex potential energy surfaces \newi{and higher dimensionalities}) need to be performed in order to pinpoint the advantages (and disadvantages) of using TTNS over MPS.

Additionally, we presented a simple procedure how to optimize (``disentangle'') a tree by randomly swapping and moving dimensions between neighboring tensors. 
As proof of concept, we showed how to optimize the tree for \ce{CH3CN}, reducing the number of required parameters by around $\unit[5]{\%}$ to $\unit[10]\%$, compared to a tree based on a \emph{very good} initial guess.
We remark that this optimization could also be performed separately for each eigenstate or even \emph{during} a time propagation, where, after some time steps, the tree changes in order to reduce the required bond dimensions.

Here, we presented only the basic methodology of TTNS and there are many possible improvements.
In particular, almost all sophisticated advances in the DMRG can straightforwardly be implemented for TTNS.
For example, there has been much work on improving excited state calculation.
Another way of improvement is the usage of dynamical or adaptive pruning (which is similar to selected configuration interaction) to further decrease the required number of parameters.
Previous studies of dynamical pruning in combination with the MCTDH algorithm are very promising.\cite{dp_wodraszka_2017,dpmctdh_tannor_2017,kohler_2019a}
At any rate, we plan to apply TTNS and MPS to larger systems with more complicated potential energy surfaces.

Furthermore, DMRG-based algorithms for time evolution\cite{haegeman_2016,chan_2018,baiardi_2019a,kurashige_2018,hubig_2019} 
can straightforwardly be applied to TTNS.\cite{schroder_2019,bauernfeind_2019}
It will be very interesting to see how they compare to the MCTDH-based algorithms. Also, we believe that the diagrammatic notation used in the DMRG community and in this work will highlight new facets of established MCTDH methodology. This will lead to better understanding and maybe even lead to further improvements that enable the computation of challenging systems.

\begin{acknowledgments}
The author is thankful to G.~K.~Chan, K.~Gunst and R.~Haghshenas for helpful discussions.
He acknowledges support from the German Research Foundation (DFG) via grant LA 4442/1-1.
\end{acknowledgments}

\end{document}